\documentclass[a4paper,twocolumn,11pt,accepted=2026-01-02]{quantumarticle}
\pdfoutput=1

\usepackage{epsf}
\usepackage{amsmath, amssymb, graphicx}
\makeatletter
\renewcommand*\env@matrix[1][*\c@MaxMatrixCols c]{%
  \hskip -\arraycolsep
  \let\@ifnextchar\new@ifnextchar
  \array{#1}}
\makeatother
\usepackage[numbers,sort&compress]{natbib}

\usepackage{hyperref}
\hypersetup{
    colorlinks=true,
    linkcolor=blue,
    citecolor=blue,
    urlcolor=blue,
    pdfborder={0 0 0}
}

\usepackage[T1]{fontenc}
\usepackage[font=small,labelfont=bf]{caption}
\usepackage{physics}
\usepackage{verbatim}
\usepackage{subcaption}
\usepackage{mathtools}
\usepackage[normalem]{ulem}
\usepackage{placeins}
\usepackage[usenames,dvipsnames]{xcolor}
\definecolor{burgundy}{rgb}{0.565,0.0,0.125}
\DeclareUnicodeCharacter{2009}{\,}
\usepackage{todonotes}
\usepackage{afterpage}

\def\be{\begin{equation}}
\def\ee{\end{equation}}
\def\bea{\begin{eqnarray}}
\def\eea{\end{eqnarray}}
\def\nn{\nonumber}

\def\ra{\rangle}

\def\sI{\mathcal{I}}
\def\sJ{J}
\def\Id{{\mathbb{I}}}

\newcommand{\CZ}{\mathrm{CZ}}

\newcommand{\beq}{\begin{eqnarray}}
\newcommand{\eeq}{\end{eqnarray}}
\def\OMIT#1{{}}
\newcommand{\lsim}{\mathrel{\rlap{\lower4pt\hbox{\hskip1pt$\sim$}}
    \raise1pt\hbox{$<$}}}         
\newcommand{\gsim}{\mathrel{\rlap{\lower4pt\hbox{\hskip1pt$\sim$}}
    \raise1pt\hbox{$>$}}}         
\newcommand{\dt}[1]{{\textcolor{blue}{DT: #1}}}
\def\AS{\textcolor{red}}

\begin{document}

\title{Lower bounds on bipartite entanglement in noisy graph states}

\author{Aqil Sajjad}
\affiliation{James C. Wyant College of Optical Sciences, University of Arizona, Tucson, AZ 85721, USA}
\author{Eneet Kaur}
\affiliation{James C. Wyant College of Optical Sciences, University of Arizona, Tucson, AZ 85721, USA}
\affiliation{Cisco Quantum Lab, Los Angeles, USA}
\author{Kenneth Goodenough}
\affiliation{College of Information and Computer Sciences, University of Massachusetts Amherst, MA 01002, USA}
\author{Don Towsley}
\affiliation{College of Information and Computer Sciences, University of Massachusetts Amherst, MA 01002, USA}
\author{Saikat Guha}
\affiliation{James C. Wyant College of Optical Sciences, University of Arizona, Tucson, AZ 85721, USA}

\begin{abstract}
Graph states are a key resource for a number of applications in quantum communication and computing. Due to the inherent noise in noisy intermediate-scale quantum (NISQ) era devices, it is important to understand the effects noise has on the usefulness of graph states. We consider a noise model where the initial qubits, prepared in $\ket{+}$ states, undergo depolarizing noise before the application of the $\CZ$ operations that generate edges between qubits situated at the nodes of the resulting graph state. For this model we develop a method for calculating the coherent information---a lower bound on the rate at which entanglement can be distilled, across a bipartition of the graph state. We also identify some patterns on how adding more nodes or edges affects the bipartite distillable entanglement. As an application, we find a family of graph states that  maintain a strictly positive coherent information for any amount of (non-maximal) depolarizing noise.
\end{abstract}

\maketitle

\section{Introduction}
A quantum network distributes quantum states amongst remote parties, who can then use the states as a resource for numerous applications, exceeding what can be done with purely classical resources~\cite{wehner2018quantum}. A relevant class of states here are {\em graph states}. These states are simple to describe, yet are key resources for various quantum applications, especially for photonic implementations. Graph states have found applications ranging from cryptography, computation \cite{PhysRevA.52.R2493,FBQC,Kitaev2003,Briegel}, and communication \cite{RevModPhys.83.33,Azuma2015} to quantum sensing \cite{Damian2020}.

Graph states, though useful, are hard to generate and distribute over a quantum network. This has motivated a significant amount of research, from tracking noisy graph states~\cite{MorRuiz2023}, to the distribution of graph states~\cite{mor2023noisy, Fischer2021}, to extraction of smaller graph states from noisy ones~\cite{szymanski2024useful, vairogs2024localizing, frantzeskakis2023extracting}, to performance of metrology under realistic noise~\cite{Damian2020}.
In this work, we analyse how noise affects entanglement across a bipartition of a graph state. In particular, we consider both depolarizing noise during the preparation of the initial qubits, and dephasing noise after the state has been distributed. To quantify the resultant state, we calculate the entropies of the total system and its two subsystems. This allows us to calculate several quantities of interest, such as the mutual information~\cite{schumacher2006quantum, berta2015renyi} or the conditional mutual information~\cite{wilde2011classical, fawzi2015quantum, brandao2015quantum,kaur2017conditional}. In this work, we will focus on the coherent information as it provides a lower bound on the distillable entanglement \cite{Devetak_2005}. Thus if we have a non-zero coherent information, then we know that we have at least that amount of bipartite entanglement. 

Our method, for which we provide an algorithm,  is based on carrying out Gaussian elimination on the biadjacency matrix, which describes which of Alice's nodes are connected to those of Bob. We then perform a deeper dive and consider the rank of the biadjacency matrix, which gives the bipartite entanglement in the noiseless case~\cite{fattal2004entanglement}. We list all ways in which we can have rank 1 or 2 for the biadjacency matrix, and consider the effect of noise in these cases.
We derive properties that allow us to identify such rank 1 and 2 graph states that yield high coherent information in the presence of noise. Furthermore, we also identify patterns as to how to divide the qubits amongst the two parties to increase the coherent information.
Based on these insights, we find a family of graph states  which we observe   numerically to have positive coherent information for any amount of (non-maximal) depolarizing noise.
These include star graphs, which have rank 1 for the biadjacency matrix, as well as a collection of higher rank graphs. Preparation of such states can then provide resilience against noise during distribution of states through quantum networks.

In what follows we first detail our model and some of the basics of graph states in Sec.~\ref{sec:background}. We then move on to the calculation of the coherent information for pure (Sec.~\ref{sec:coh_perfect}) and noisy (Sec.~\ref{sec:coh_noisy}) graph states, before comparing our results in Sec.~\ref{sec:results}. Finally, we end with some concluding remarks in Sec.~\ref{sec:conclusion}.

\section{Background and model}\label{sec:background}
In this section, we give relevant definitions and results from existing literature as background material.

\subsection{Introduction to graph states and stabilizers}
Let us first set some graph-theoretic notation. Informally, a graph consists of a set of vertices, some pairs of which are connected by edges. An $n$-vertex graph can be specified by an $n\times n$ binary-valued {\em adjacency matrix} $G$, such that $G_{ij} = 1$ if vertices $i$ and $j$ are connected by an edge and $G_{ij} = 0$ otherwise. Note that we implicitly assume an ordering on the vertices. We can always choose a labelling such that $G$ has the following form for any bipartition $A, B$ of the vertices,
\be
G =
\begin{pmatrix}
G_A & G_{AB} \\ 
G_{AB} ^T & G_B
\end{pmatrix} \ ,
\ee
where $G_A$ and $G_B$ respectively are the adjacency matrices of the subgraphs corresponding to $A$ and $B$. The matrix $G_{AB}$ is called the {\em biadjacency matrix}. It specifies the edges connecting Alice's qubits with those owned by Bob, and is key in our description of the entanglement across the bipartition $A,~ B$.

Graph states are simple to describe, yet rich enough to capture all of the entanglement structure in stabilizer states,  see~\cite{gottesman1997stabilizer} for an introduction to the stabilizer formalism and~\cite{hein2006entanglement} for an introduction to graph states.
A graph state on $n$ qubits is described by an $n$-vertex graph. Each qubit corresponds to a vertex in the graph, and is initialized in $\ket{+}$. Then for each edge connecting two vertices in the graph, we apply a $\CZ$ operation on the corresponding pair of qubits to obtain the graph state. Note that $\CZ$ gates commute, and therefore the order in which they are applied does not matter, hence the above is well-defined.
In what follows, we will sometimes take the liberty to refer to edges and $\CZ$ operations interchangeably. Thus, if we say that two qubits are connected by an edge, it means there is a $\CZ$ operation on them. 

We can describe a graph state equivalently in terms of its so-called stabilizer group. For each qubit $i$ where $i=1\ldots n$, define a corresponding operator $S_i$ as the Pauli $X$ matrix acting on qubit $i$ and a $Z$ operator on each qubit with which it is "connected" by a $\CZ$ operation:
\be
S_i = X_i \prod_{j\in N_{v_i}} Z_j \ .
\ee
Here $N_{v_i}$ is the set of qubits corresponding to the neighborhood of vertex $i$ in the graph, i.e. the set of vertices that are connected with vertex $i$ through edges.
The operators $S_i$ form a generating set of an abelian group called the stabilizer group. The graph state is then the unique eigenstate of the entire stabilizer group with eigenvalue 1. 

\subsection{Our noise model and modified graph states}
For our noise model we consider depolarizing noise after the creation of the $n$ qubits in the $\ket{+}^{\otimes n}$ state prior to the application of the $\CZ$ gates. Note that depolarizing $\ket{+}$ states is equivalent to applying dephasing noise to them. Under such noise, each qubit remains in $\ket{+}$ with some probability $P$, and flips into $\ket{-}$ with probability $1-P$.
We should mention that since $Z$ gates commute with $\CZ$ gates, dephasing noise on the initial $\ket{+}$ qubits before the $\CZ$
 gates is equivalent to first applying the $\CZ$ operations followed by dephasing. In principle, we can have noise both before and after the $\CZ$ gates, but due to this equivalence, we can parameterize the combined effect of both with a single application of dephasing noise on each qubit before the $\CZ$ operations.

Our choice of the symmetric noise model where each qubit is affected by the same amount of dephasing noise is motivated by its practical relevance. Early-generation quantum networks are likely to involve short distances between and uniform hardware across nodes, leading to comparable noise levels for all qubits. While asymmetric noise scenarios may become relevant in next-generation networks with heterogeneous hardware, extreme cases (e.g., $P_A \ll P_B$ or $P_B \ll P_A$) seem unlikely, as both parties would aim to minimize noise through comparable technologies. That said, our framework is capable of handling asymmetric noise, which is being considered in~\cite{goodenough2024bipartite}.

In light of the above noise model, our noisy graph state will be a mixture of standard as well as \emph{modified graph states} in which a subset of the $n$ qubits are flipped from $\ket{+}$ to $\ket{-}$ before the application of the $\CZ$ gates, see for example~\cite{goyal2006purification, patil2023clifford}. The modified graph states constituting this mixture can still be individually described by the stabilizer formalism. Note that a $Z$ gate leaves a $Z$ operator in a stabilizer generator invariant, but flips an $X$ into $-X$. Consequently, for each initial qubit $i$ transformed from $\ket{+}$ to $\ket{-}$ due to the application of a $Z_i$ operator from dephasing, the corresponding stabilizer generator acquires a minus sign.

\subsection{The coherent information}
        
We want to answer the question of how \emph{useful} a given noisy graph state is. One qualitative answer is given by the rate at which Bell pairs can be distilled from asymptotically many copies of such a noisy graph state, where Alice and Bob are allowed to use local operations and one-way classical communication. This is a complicated question in general. We consider \emph{coherent information} as it provides a lower bound on the distillable entanglement~\cite{Devetak_2005}. We emphasize that our tools can be used to calculate any quantity based on the entropies of the total and subsystems, such as the mutual information.

Consider a mixed state $\rho_{AB}$ shared by Alice and Bob. The coherent information is given by
\be
\sI_B = H(\rho_B) - H(\rho_{AB})\ ,
\label{coherent-information-definition}
\ee
where $\rho_B$ is the reduced density matrix for Bob, and $H(\rho)$ is the Von Neumann entropy function given by
\be
H(\rho) = -\mathrm{Tr}\left(\rho \log_2\rho\right) \ .
\ee
A related quantity is the reverse coherent information \cite{Garc_a_Patr_n_2009}, defined by
\be
\sI_A = H(\rho_A) - H(\rho_{AB})
\label{reverse-coherent-information-definition}\ .
\ee
There is some arbitrariness as to which one we call which, and it is possible to confuse the two.
For ease of exposition, we will at times refer to $\sI_A$ ($\sI_B$) as ``Alice's coherent information'' (``Bob's coherent information''), ``the coherent information on Alice's side'' (``coherent information on Bob's side''), or simply ``coherent information $\sI_A$'' (``coherent information $\sI_B$'). Furthermore, we will refer to the general quantity as coherent information when we are not spelling out whether we are talking about $\sI_A$ or $\sI_B$. Note that $\max\{\sI_A,\sI_B\}$ yields the tightest lower bound on distillable entanglement.

\subsection{Coherent information for a noiseless graph state}

Since a noiseless graph state is a pure state, the Von Neumann entropy of the full state is simply zero. Therefore, the coherent informations $\sI_A$ and $\sI_B$ equal the bipartite entanglement entropy i.e. the Von Neumann entropy of either subsystem which are known to equal each other~\cite{wilde2011classical}.
It has been shown in ~\cite{fattal2004entanglement} that the entanglement entropy for a bipartite pure graph state is given by the rank of the biadjacency matrix. We will have more to say on this connection with the rank in the coming sections.

\subsection{The pure state density matrix in terms of stabilizer generators}

The density operator associated with a stabilizer state can be expressed in terms of its stabilizer generators as
\be
\rho = \frac{1}{2^n} \prod_{k=1}^n \left(\mathbb{I} +S_k\right),
\ee
where $\mathbb{I}$ is the identity operator on the $n$-qubit Hilbert space. We will refer to each term  $\left(\mathbb{I} +S_k\right)$ as a \emph{bracket}.
The reason why the above formula works is simple. Each bracket $\left(\mathbb{I} +S_k\right)$
gives zero when acting on eigenstates of $S_k$ with eigenvalue $-1$. It returns the eigenstates of $S_k$ that have eigenvalue 1 but multiplies them by 2. The product of all such brackets selects the unique state of all the stabilizer generators with eigenvalue 1, which is normalized by the $1/{2^n}$ factor. Note that the commutativity of the stabilizer group is crucial for this argument, also ensuring that the product is well-defined without specifying an ordering.

\subsection{The reduced subsystem density operator in terms of stabilizer generators}
\label{tracing-out-section}

In the same spirit, the reduced density operator on a subsystem of qubits is given by a similar product. That is, it involves the generators of the subgroup of the stabilizers that act non-trivially only on the qubits belonging to that subsystem (i.e. they have the identity on all the qubits not in the subsystem). 
Here and in what follows, we implicitly prune the generators from the identity operators acting on the system that gets traced out.
Specifically, let us say Alice has $n_A$ qubits and Bob has $n_B$ qubits, and $S_A$ and $S_B$ are the subgroups of the stabilizer group that act non-trivially only on subsystems $A$ and $B$ (and trivially on systems $B$ and $A$), respectively.
Then the reduced density matrices for subsystems $A$ and $B$ are
\bea
\rho_A=&\frac{1}{2^{n_A}} \prod_{k=1}^{K_A}
\left(\mathbb{I} +S_{A_k}\right) \nn \ , \,{\text{and}}\nonumber \\
\rho_B=&\hspace{-8.85mm}\frac{1}{2^{n_B}} \prod_{k=1}^{K_B}
\left(\mathbb{I} +S_{B_k}\right)
\label{reduced-density-matrix}
\eea
where the $S_{A_k}$ ($S_{B_k}$) are generators for $S_A$ ($S_B$), and $K_A$ ($K_B$) is the (minimum) number of such generators~\cite{hein2004multiparty}. We will see in the next section that the cardinality $K_A$ ($K_B$) of $S_A$ ($S_B$) is closely related to the binary rank of the biadjacency matrix 
\bea
K_A &\equiv |S_A| &= n_A -\textrm{rank}(G_{AB}) \nn \ , \\
K_B &\equiv |S_B| &= n_B -\textrm{rank}(G_{AB}) \ ,
\label{cardinality-rank-relationship}
\eea
where $\textrm{rank}(G_{AB})$ is the binary rank.

Formula (\ref{reduced-density-matrix}) is satisfied by all stabilizer states. Therefore, it also applies to our \emph{modified graph states} in which some or all qubits flip to $\ket{-}$ before the $\CZ$ gates. Also note in (\ref{reduced-density-matrix}) that since each bracket has non-zero eigenvalue $2$, the reduced state $\rho_A$ ($\rho_B$) is an equal probability mixture of $2^{n_A -K_A}$ ($2^{n_B -K_B}$) pure states on the $n_A$ ($n_B$) qubits in the subsystem. If $K_A$ ($K_B$) is zero, then the reduced state is simply the maximally mixed state on the reduced subsystem. 

\section{Tracing out a subsystem in a pure graph state}
\label{sec:coh_perfect}

We now present a procedure for obtaining a generating set for $S_A$ or $S_B$ in a graph state. This can then be used to obtain the reduced density matrix $\rho_A$ or $\rho_B$ via \eqref{reduced-density-matrix}. We will also show how to adapt this method for {\em modified graph states}---graph-like states where a subset of the $n$ qubits are initialized in the $\ket{+}$ state and the remaining in the $\ket{-}$ state before CZs are applied along the edges of the graph. Thus in this section, we will show how to obtain the reduced density operators for strictly pure modified graph states. Then in the next section, we will apply these results and methods to the case of our noisy graph states. We make use of the fact that our noise model yields a convex mixture of pure modified graph states, of which we can calculate the associated probabilities, from which we calculate the coherent information.

\subsection{Obtaining generators for $S_A$ and $S_B$ through Gaussian elimination}

We focus now on describing the procedure for 
obtaining the generators of $S_A$, understanding that an identical process exists for (the generators of) $S_B$. To obtain the generators $\{S_{A_k}\}$, we multiply the various stabilizer generators to get them in a form where as many of them as possible act trivially on subsystem $B$. We do this by writing the stabilizer generators in terms of the biadjacency matrix $G_{AB}$ (see Sec.~\ref{sec:background}).

First, note that the stabilizer generators have a simpler form when there are no local edges between the qubits of the same party, and the only edges are across the bipartition. In fact, we can remove all local edges by applying $\CZ$ gates without changing the coherent information. This is because a $\CZ$ gate on two qubits of the same party is a local unitary operation within the same subsystem and does not change any of the Von Neumann entropies. In the absence of local edges, the stabilizer generators associated with Alice's qubits now only have $X$ operators acting on her qubits and only $Z$ operators on Bob's qubits, and vice versa.

With no local edges, we can describe the stabilizer generators in terms of the following two matrices:
\bea
G_{\rm ext, A} &\equiv \begin{pmatrix}
I_A & G_{AB} 
\end{pmatrix} \nn\ , \\
G_{\rm ext, B} &\equiv \begin{pmatrix}
G_{AB} ^T & I_B
\end{pmatrix} \ ,
\label{extended-biadjacency-matrices}
\eea
where $I_A$ and $I_B$ are $n_A\times n_A$ and $n_B\times n_B$ identity matrices, respectively.
The first $n_A$ stabilizer generators are described by $G_{\rm ext, A}$ with $X$ operators corresponding to the 1 entries in $I_A$ and $Z$ operators to 1s in the $G_{AB}$ block.
This is because if we have an edge between Alice's $i$'th qubit and Bob's $j$'th qubit, then the $ij$ entry of the biadjacency matrix $G_{AB}$ is 1, and we have a $Z$ on Bob's $j$'th qubit in the $i$'th stabilizer generator. On the other hand, if there is no edge then there is a zero entry in the biadjacency matrix, corresponding to the identity on Bob's $j$'th qubit. In the same way, we can express the remaining $n_B$ generators in terms of the rows of $G_{\rm ext, B}$ with $X$ operators described by $I_B$ and $Z$'s by $G_{AB}^T$.

A (mod 2) addition of two rows of $G_{\rm ext, A}$ describes the product of the two corresponding stabilizer generators. Specifically, each 1 (0) in the first $n_A$ entries of the resulting row represents an $X$ (identity) on the corresponding qubit in the new stabilizer generator obtained through the multiplication of the two original generators. In the remaining $n_B$ entries of the resulting row, 1 (0) represents a $Z$ (identity) on the corresponding qubit.

In order to obtain the generators for $S_A$, we perform a set of row operations on $G_{\rm ext, A}$ to reduce the $G_{AB}$ block to its row echelon form. We carry along the other $I_A$ block, and perform the same row operations on it.
When we are finished, each zero row in the row echelon form of $G_{AB}$ corresponds to a stabilizer generator that acts trivially on Bob's subsystem.
It therefore represents a generator of the subgroup $S_A$, and is described by the corresponding row in the final form of the $I_A$ block with an $X$ for each one entry and the identity for each zero.
Thus if a zero row in the $G_{AB}$ block is obtained by adding rows 1, 2 and 4, then we obtain $X_1 X_2 X_4$ as the corresponding generator in $S_A$.
A similar procedure gives the generators for subsystem $B$ by reducing $G_{AB} ^T$ in $G_{\rm ext, B}$ to its row echelon form and reading off the row entries of the final form of the $I_B$ block for the $X$ operators in the generators of $S_B$.

This procedure also makes the connection (\ref{cardinality-rank-relationship}) between the rank of the biadjacency matrix and the cardinalities $K_A$ and $K_B$ clear. The number of uneliminated rows of $G_{AB}$ equals the rank and is given by $n_A-K_A$. Likewise, on Bob's side, the number of uneliminated rows of $G_{AB} ^T$  (i.e. the uneliminated columns of $G_{AB}$ in its column echelon form) is equal to $n_B-K_B$.

Lastly, we should mention that the procedure for obtaining the generators of $S_A$ and $S_B$ described above also holds if some of the qubits are initialized in $|-\rangle$ instead of $|+\rangle$, but now the corresponding $X$ operators in the stabilizer generators change to $-X$, introducing possible negative signs in some of the generators. Therefore, new generators involving tensor products of $X$ operators on different qubits will have a $+$ ($-$) sign if an even (odd) number of the original stabilizers multiplied together have an even (odd) number of minus signs.

\subsection{An example}

\begin{figure}
    \centering
\includegraphics[scale = .3]{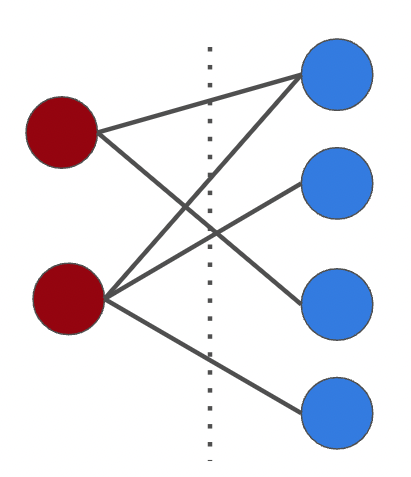}
    \caption{In this figure, the red qubits belong to Alice and the blue qubits belong to Bob. Alice has qubits 1 and 2, and Bob has the remaining four. The graph has edges $13$, $15$, $23$, $24$ and $26$. }
    \label{fig:example_2}
\end{figure}

Let us illustrate this with an example. Consider the 6 qubit graph in which Alice has qubits 1 and 2, and Bob has the remaining four. The graph has edges $13$, $15$, $23$, $24$ and $26$. We illustrate this graph state in Fig~\ref{fig:example_2}. This graph has the biadjacency matrix
\be
G_{AB} = \begin{pmatrix}
1 & 0 & 1 & 0 \\
1 & 1 & 0 & 1 \end{pmatrix}\ ,
\label{6-qubit-example-adjacency}
\ee
Furthermore, the stabilizer generators are given by
\begin{gather}
\begin{pmatrix}
    \begin{tabular}{cc|cccc}
$X$ & $I$ & $Z$ & $I$ & $Z$ & $I$\\
$I$ & $X$ & $Z$ & $Z$ & $I$ & $Z$\\
\hline
$Z$ & $Z$ & $X$ & $I$ & $I$ & $I$\\
$I$ & $Z$ & $I$ & $X$ & $I$ & $I$\\
$Z$ & $I$ & $I$ & $I$ & $X$ & $I$\\
$I$ & $Z$ & $I$ & $I$ & $I$ & $X$\\
\end{tabular}
\end{pmatrix}
\end{gather}
Here the horizontal line has been included to separate the stabilizer generators corresponding to Alice and Bob's qubits, and the vertical line separates the Pauli operators acting on Alice and Bob's subsystems.

Now, it is clear that we cannot obtain a zero row through any row additions in the biadjacency matrix $G_{AB}$. Consequently we cannot multiply the first two stabilizer generators to obtain a product that acts non-trivially only on Alice's subsystem (i.e. have identities on all of Bob's qubits). The last four stabilizer generators are already excluded from the possibility of giving such a product because of the $X$ operators on different qubits in Bob's subsystem. 
Therefore the stabilizer subgroup $S_A$ acting trivially on Bob's subsystem is just the identity operator, and Alice's reduced state is the fully mixed state.

As for Bob's subsystem, we obtain from (\ref{extended-biadjacency-matrices})
\begin{align}
G_{\rm ext, B} & = \begin{pmatrix}
     \begin{array}{c|c}
G_{AB} ^T & I_B
\end{array}
\end{pmatrix} \ \nonumber
, \\
 & = \begin{pmatrix}
    \begin{array}{cc|cccc}
1 & 1 &  1 & 0 & 0 & 0 \\
0 & 1 &  0 & 1 & 0 & 0 \\
1 & 0 & 0 & 0 & 1 & 0 \\
0 & 1 & 0 & 0 & 0 & 1 \\
\end{array}
\end{pmatrix} \ .
\end{align}

Now, we carry out row operations to transform the $G_{AB} ^T$ block to its row echelon form (which is equivalent to  converting $G_{AB}$ to its column echelon form). 
These row operations are also performed on the $I_B$ block. 
Adding the first three rows gives a zero row in the left block, so we replace the third row by the sum of rows 1, 2 and 3. Likewise, adding rows 2 and 4 also gives a zero row, so we replace row 4 by the sum of rows 2 and 4.
We thus obtain the reduced form of $G_{\rm ext, B}$,

\be
G_{\rm ext, B} '
= \begin{pmatrix}
    \begin{array}{cc|cccc}
1 & 1 &  1 & 0 & 0 & 0 \\
0 & 1 &  0 & 1 & 0 & 0 \\
0 & 0 & 1 & 1 & 1 & 0 \\
0 & 0 &  0 & 1 & 0 & 1 \\
\end{array}
\end{pmatrix} \ .
\ee

We are left with two zero rows in the left $4\times 2$ block, so there are $K_B=2$ generators of $S_B$, which are given by $X$ operators corresponding to $1$s in the right block. Thus row 3 corresponds to the operator $X_3 X_4 X_5$, and row 4 to $X_4 X_6$. We can confirm this by carrying out the same sequence of multiplications on the last four stabilizer generators as the row operations we performed on $G_{\rm ext, B}$.
The above are the generators of $S_B$ if all of Bob's qubits are initialized in $|+\rangle$, and we obtain the following reduced density matrices
\bea
\rho_A &= &\frac{1}{4}{\mathbb{I}} \ ,  \\ 
\rho_B &= &\frac{1}{16} \left({\mathbb{I}} + X_3 X_4 X_5\right) \left({\mathbb{I}} + X_4 X_6\right),
\eea
If some of these qubits are initialized in $|-\rangle$ instead of $\ket{+}$, then the only thing that changes is that the $X$ operators on those qubits have minus signs. Thus if an odd number of $X$s in a bracket acquire a minus sign, the $+$ sign flips to $-$. We will use this idea in the next section, where the noisy state is a convex combination of modified graph states.

\subsection{Useful notation for modified graph states} \label{useful-notation-section}

We now introduce some notation, convenient for describing the stabilizers of modified graph states, and useful in the next section when we consider the noisy case. The $K_A\times n_A$ lower left corner block of the final form of $G_{\rm ext, A}$ (obtained after the row reductions) describes the generators of $S_A$; we denote it by $\sJ_A$.
We can therefore write these generators as
\be
S_{A_k} = \prod_{l=1}^{n_A} \left(s_l X_l\right) ^{\sJ_{A, kl}}\ ,
\label{generators-in-terms-of-sJ}
\ee
where $s_l$ equals $1$ ($-1$) if the original stabilizer involving $X_l$ has a $+$ ($-$) sign (due to the corresponding qubit being initialized in $|+\ra$ or $|-\ra$). Here, $\sJ_{A, kl}$ is the entry of $\sJ_A$ labeled by indices $k$ and $l$.

Alternatively, we can write the generators as
\be
S_{A_k} = (-1)^{m_{A, k}} \prod_{l=1}^{n_A} X_l ^{\sJ_{A, kl}}\ ,
\label{generators-in-terms-of-m_k}
\ee
where $m_{A, k}$ is zero (one) if we have an even (odd) number of $-$ signs in the $k$-th bracket for subsystem $A$:
\be
m_{A, k} \equiv \frac{1}{2}\left[1 -\prod_{l=1}^{n_A} \left(s_l \right) ^{\sJ_{A, kl}}\right]\ .
\ee
Henceforth $\vec m_A$ refers to the entire collection of $m_{A, k}$ values specifying the whole bracket configuration for all the generators of $S_A$. 

The reduced density matrix $\rho_{A, \vec m_A}$ for  Alice's subsystem for a given bracket sign configuration $\vec m_A$ can now be written in terms of these generators by using \eqref{reduced-density-matrix},
\be
\rho_{A, \vec m_A} = \frac{1}{2^{n_A}}\ 
\prod_{k=1}^{K_A} \left({\mathbb{I}} + (-1)^{m_{A, k}}\  \prod_{l=1}^{n_A} X_l ^{\sJ_{A, kl}}\right)\ ,
\label{eq:gen_sigma}
\ee
where ${\vec m_A} \equiv (m_1, m_2, \ldots, m_{K_A})$
describes the signs in all the brackets.

The prefactor ${1}/{2^{n_A}}$ along with the product of the $K_A$ brackets yields the non-zero eigenvalue $2^{K_A -n_A}$ with a multiplicity of
$2^{n_A -K_A}$ for each $\vec m_A$. The remaining
$2^{n_A} - 2^{n_A -K_A}$ 
eigenvalues are all zero.

\section{Coherent information for a noisy graph state}\label{sec:coh_noisy}

Having spelled out what happens for modified graph states, we are now ready to consider noisy states, i.e., convex combinations of modified graph states.

\subsection{The full state entropy}

With each qubit starting in $|+\ra$ with probability $P$ and $|-\ra$ with probability $1-P$, the starting Von Neumann entropy of each qubit is simply the binary entropy function
\be
H_2(P) \equiv -P\log_2(P) \,-\,(1-P)\log_2(1-P).
\label{H_2-definition}
\ee
Since we have $n$ qubits, the total entropy is $n H_2(P)$. The $\CZ$ operations associated with the edges in the graph are unitary operations which do not change the entropy. Therefore, we conclude that the Von Neumann entropy of the full system is always
\be
H(\rho_{AB})
= n H_2(P)\ .
\label{Vonn-Neumann-entropy-full-noisy-system}
\ee

\subsection{The subsystem entropy}
Recall the form for the reduced density operator $\rho_{A, \vec m_A}$ from \eqref{eq:gen_sigma} when the full system is in a standard or (pure) modified graph state (in which some or all the qubits are initialized in $|-\ra$).
We now consider noisy mixtures of such states and describe the entropy of the reduced subsystem.
The reduced state for subsystem $A$ for our noisy system is the weighted sum over all the bracket configurations
\begin{equation}
\rho_A = \sum_{{\vec m_A}} w_{\vec m_A} \rho_{A, \vec m_A}\ ,
\label{weighted-sum-reduced-matrix}
\end{equation}
where each weight $w_{\vec m_A}$ is the total probability of all initial qubit configurations that yield $\vec m_A$.
Recall from Section \ref{useful-notation-section} that only the qubits corresponding to the non-zero columns of $\sJ_A$ determine $\vec m_A$.
Therefore, we only need to sum over the initial configurations of these relevant qubits, since for the other qubits, the probabilities $P$ and $1-P$ for initialization in $|+\ra$ and $|-\ra$ simply sum to one.
Suppose we have $\nu_A$ non-zero columns of $\sJ_A$. The probability of a given initial qubit configuration involving the corresponding $\nu_A$ qubits is
$P^{\nu_A-\nu_-} (1-P)^{\nu_-}$,
where $\nu_-$ is the number of qubits initialized in $|-\ra$, so the remaining $\nu_A-\nu_-$ start in $|+\ra$.
The weight $w_{\vec m_A}$ is the sum of all such probabilities for the configurations giving $\vec m_A$.
In Section \ref{sec:algorithm}, we present an algorithm for calculating the coherent information. It involves a procedure for calculating these weights by iteratively adding over all the qubit configurations.

To calculate the coherent information $\sI_A$, we need to obtain the von Neumann entropy of the density operator (\ref{weighted-sum-reduced-matrix}). Recall from section (\ref{tracing-out-section})
that for each $\vec m_A$, $\rho_{A, \vec m_A}$ has $2^{n_A -K_A}$ identical non-zero eigenvalues, each being $2^{K_A -n_A}$.

It is also easy to see that each of the $\rho_{A, \vec m_A}$ density operators are diagonal in the $|\pm\rangle^{\otimes n_A}$ basis, and have mutually-orthogonal (non-overlapping) supports. The (non-zero) eigenvalues of $\rho_A$ are given by $\left\{\frac{1}{2^{n_A-K_A}}w_{\vec m_A}\right\}$ with multiplicity $2^{n_A-K_A}$. Writing $H\left(\left\{w_{\vec m_A}\right\}\right) = -\sum_{{\vec m_A}=0}^{2^{K_A}-1}w_{\vec m_A}\log_2 w_{\vec m_A}$, we find
\be
H(\rho_A) = H\left(\left\{w_{A, \vec m_A}\right\}\right) 
+ n_A-K_A \ .
\ee
We are now in a position to calculate the coherent information of any graph state with a given bipartition. We present an algorithm for calculating the coherent information by simple brute force in Appendix~\ref{sec:algorithm}.  The algorithm iteratively loops through all of the initial configurations of the qubits, and adds up all their contributions to the weights $w_{A, \vec m_A}$.

We can gain insight by calculating the weights and the coherent information for some simple cases. In what follows, we detail all cases where $G_{AB}$ has rank 1 or 2. It turns out that the former corresponds to fully connected graphs, and the latter can be categorized into two types of subsystems for each party. 

\subsection{The rank 1 i.e., $K_A=n_A-1$ case}
\label{rank1-section}

\begin{figure}
    \centering
    \includegraphics[scale=0.3]{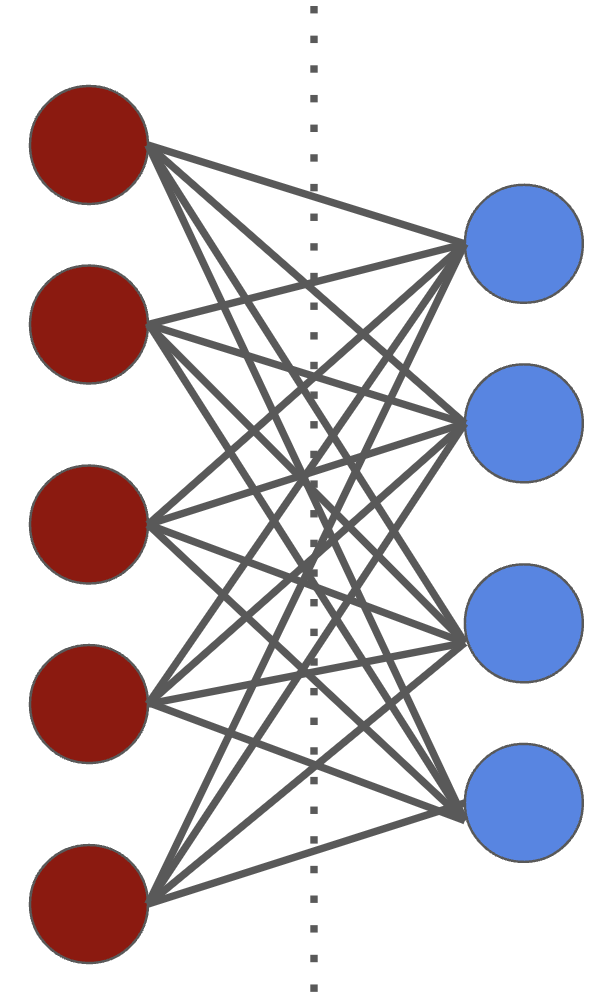}
    \caption{Red qubits belong to Alice and blue qubits belong to Bob. This is an example of a rank 1 case with $n_A = 5$ and $n_B = 4$.}
    \label{fig:rank1}
\end{figure}

The rank 1 case corresponds to all of Alice's qubits being connected to all of Bob's qubits through edges i.e., $\CZ$ operations, see Fig~\ref{fig:rank1} for an example. Every graph whose biadjacency matrix $G_{AB}$ has rank 1 can be reduced to this case, along with one or more vertices of one party that are not connected to any vertex belonging to the other party.

The above observation is due to the fact that a rank 1 biadjacency matrix requires all non-zero rows (and all non-zero columns) to be identical.  Consequently, we can take one of the non-zero rows and add it to all of the other identical rows to obtain zero rows, leaving us with just one non-zero row. The same holds for the columns, and therefore, the entries of the biadjacency matrix $G_{AB}$ having ones describe a fully connected sub-graph. Zero entries correspond to vertices that have no edge across the bipartition, and therefore do not participate in sharing entanglement across the bipartition. We can ignore such qubits and focus on fully connected graphs whose biadjacency matrices have entries that are all ones.

The reduced density matrix for subsystem $A$ will be a weighted sum of the reduced states for each state in a convex decomposition of $\rho_{AB}$. Since $\rho_{AB}$ is diagonal in the graph state basis, each such state is described by a modified graph state.
\begin{align}
\rho_{A, \vec m_A} 
  &= \frac{1}{2^{n_A}} 
     \prod_{j=2}^{n_A} 
     \Big(
        \Id + (-1)^{m_{A, j-1}} X_{1} X_{j}
     \Big), \nonumber \\
  &\quad \vec{m}_A \in \{0,1\}^{K_A}.
\end{align}
Here $m_{A, j} = 0$ if both qubits $1$ and $j-1$ are initialized in the $|+\ra$ or $|-\ra$ state before applying the $\CZ$ operations, and $m_{A, j} =1$ otherwise.

To do the combinatorics, we divide the terms into two categories: one in which the first qubit belonging to subsystem $A$ is initiated in $|+\ra$ and the other in which it is initiated in $|-\ra$.
If it is in $|+\ra$, then any stabilizer generator associated with another qubit that is initiated in $|+\ra$ has a $+$ sign, and a $-$ sign for any qubit (other than qubit 1) starting in $|-\ra$.
Likewise, if qubit one starts in $|-\ra$, then we get a $+$ sign for any other qubit initiated in $|-\ra$, and a $-$ sign otherwise.
Thus each combination of the different signs associated with the stabilizer generators of the reduced subsystem appears twice, one associated with qubit one in $|+\ra$, and another where it is in $|-\ra$.

All in all, we obtain the reduced density matrix for subsystem $A$
\begin{align}
\rho_A = &~\frac{1}{2^{n_A}} \nonumber,
\sum_{ \mathclap{\substack{\\ \vec m_A \in \{0,1\}^{n_A-1} }}} \,
w_{\vec m_A}\prod_{j=2}^{n_A} 
\left(\Id+ (-1)^{m_{A, j-1}} X_{1} X_{j}\right), \nn \\
w_{\vec m_A} =&~\left[P^{n_A-i} (1-P)^i +P^i (1-P)^{n_A -i}\right] \nn , \\
i = &~\sum_{j=2}^{n_A} m_{A, j-1} .\nonumber
\label{individual-term-all-edges-reduced-density-matrix}
\end{align}
Here $i$ is the number of brackets (or stabilizer generators) with negative signs i.e., $m_{A, j-1} =1$.
The term $P^{n_A-i} (1-P)^i$ arises from $X_1$ having a positive sign and $i$ of the $X$s on the remaining qubits having negative signs.
Likewise, the $P^i (1-P)^{n_A -i}$ term involves a minus sign for $X_1$
and $i$ of the $X$s on all the other qubits having positive signs.
Each bracket $(1+(-1)^{m_{A, j-1}} X_{1} X_{j})$ has eigenvalues 0 and 2 both with multiplicity 2.

The product $\prod_{j=2}^{n_A} \left(1+(-1)^{m_{A, j-1}} X_{i} X_{j}\right)$
has two eigenstates with eigenvalues $2^{n_A -1}$.
Combining this with the ${1}/{2^{n_A}}$ yields eigenvalues ${1}/{2}$ for both eigenstates.
From the combinatorics, we see that each of these has a probability factor
$\overline{P}_i \equiv \left[P^{n_A-i} (1-P)^i +P^i (1-P)^{n_A -i}\right]$, where $i=0\ldots n_A-1$ is the number of brackets with negative signs.
We thus end up with eigenvalues
${\overline{P}_i}/{2}$
with multiplicity $2\binom{n_A-1}{i}$.
The Von Neumann entropy of the reduced subsystem is
 \begin{gather}
H(\rho_A)
= f_1(n_A, P)\\
\equiv 
 1-\sum_{i=0} ^{n_A-1}
 \binom{n_A-1}{i}
\overline{P}_i\log(\overline{P}_i).
\label{rank1-entropy}
\end{gather}
As such, the coherent information $\sI_A$ becomes
\begin{align}
\sI_A =\,&f_1(n_A, P) -n H_2(P) \nn \\
=\,&f_1(n_A, P) -n_A H_2(P) -n_B H_2(P),
\label{rank1-coherent-info}
\end{align}
where $f_1(n_A, P)$ was defined in \eqref{rank1-entropy} and the second line shows the dependence on  $n_A$ and $n_B$.

\begin{figure}
    \includegraphics[width = .5\textwidth,trim={1.15cm 0.2cm 0 0},clip]{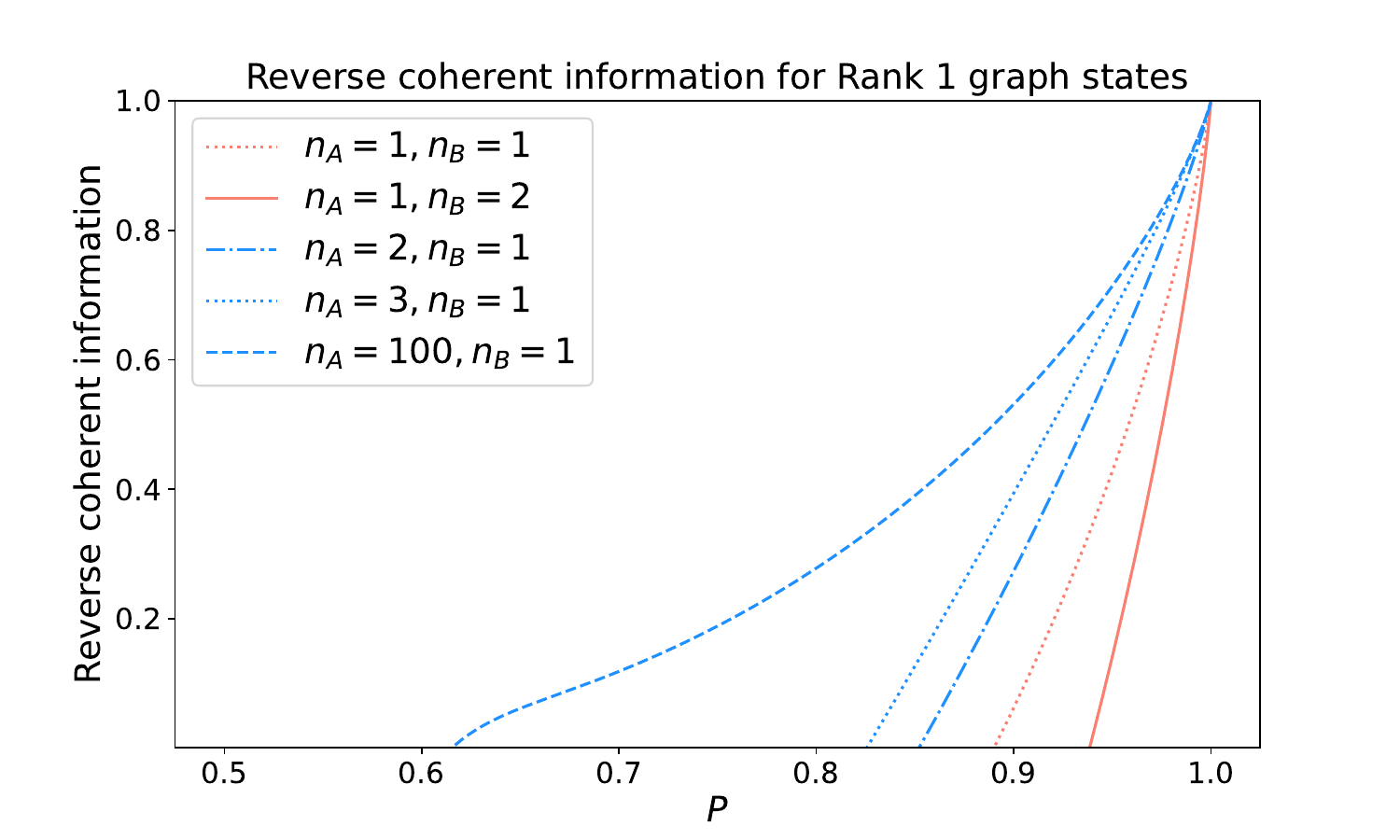}
    \caption{This plot illustrates the coherent information, $\mathcal{I}_A$, for various rank-1 cases as described in Eq.~\ref{rank1-coherent-info}. Here, $P$ represents the noise parameter, $n_A$ denotes the number of qubits held by Alice, and $n_B$ represents the number of qubits held by Bob.}
    \label{fig:rank_1_plot}
\end{figure}

For $P=0$ or $P=1$, which correspond to the noiseless case, we obtain $\sI_A=1$ (i.e. equal to the rank as expected) for all values of $n_A$ and $n_B$. For $P\neq 0$ and $P\neq 1$, we summarize the behavior as follows:
\begin{enumerate}
\item Given $n_A$, $\sI_A$ is a decreasing function of $n_B$. Therefore, $n_B = 1$ maximizes $\sI_A$. This corresponds to a star graph where Bob has only one qubit connected to all of Alice's $n_A$ qubits.
\item Given $n_B$, $\sI_A$ is an increasing function of $n_A$, which Asymptotically converges from below to a constant value for a given $P\neq 0$ and $P\neq 1$. 
\item Maximizing coherent information $\sI_A$, requires $n_B=1$ and a sufficiently large $n_A$ to reach the asymptotic limit. We show in Fig.~\ref{fig:rank_1_plot} that this gives purely positive coherent information for any non-maximal value of $P$.
\item $n_A=1$ is a trivial case for which Alice's reduced density matrix is the maximally mixed state on her qubit. This gives $H(\rho_A) = f_1(1, P) =1$,
with coherent information $\sI_A = 1-n H_2(P)$.
\item When $n_A > n_B$, $\sI_A >\sI_B$ and vice versa. That is, for any fully connected graph, the side having the largest number of qubits has the highest subsystem entropy (since $f_1(n_A, P) > f_1(n_B, P)$ when $n_A>n_B$), hence the largest coherent information. 
\end{enumerate}
See Fig.~\ref{fig:rank_1_plot} for details. 

\begin{figure}
    \includegraphics[width = .5\textwidth,trim={0.5cm 0.1cm 0 0},clip]{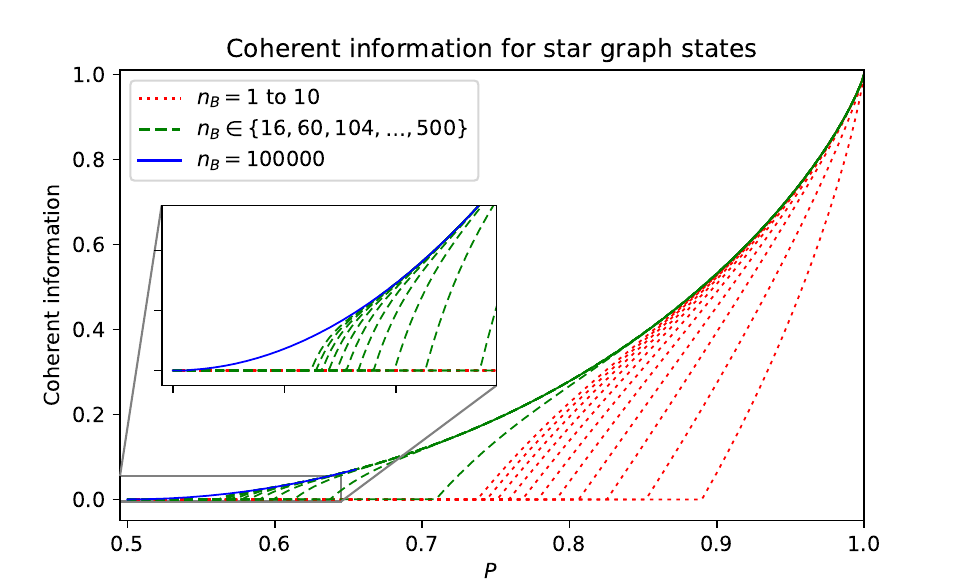}
    \caption{The coherent information of star graph states, with increasing number of qubits on Bob's side. $P$ corresponds to the noise parameter. $n_B$ represents the number of qubits held by Bob. }
    \label{fig:tolerate}
\end{figure}

Most importantly, we present numerical evidence in Fig.~\ref{fig:tolerate} that it is possible to tolerate any amount of (non-maximal) depolarizing noise by adding more qubits to one side for star graphs (or equivalently, complete graphs). In Appendix \ref{star-graph-repetition-code-appendix} we interpret this statement through the lens of a classical repetition code, allowing us to understand where the robustness of the state against noise comes from.

\subsection{The rank 2 i.e., $K_A=n_A-2$ case}
\label{rank2-structure}

Assuming that $G_{AB}$ has no zero row or column, there are two ways in which $G_{AB}$ can have rank 2. We will call them type 1 and type 2 subsystems. We will discuss these in terms of the row rank but the same classification also holds for columns. However, while the row and column ranks are equal, it is possible for a matrix to be of one type in terms of its rows and another in terms of its columns, e.g.~
 \be
\begin{pmatrix}
1 & 0 & 0 \\
0 & 1 & 1 \\
1 & 1 & 1 \\
1 & 1 & 1
\end{pmatrix} \ .
\label{example-biadjacency-matrix-different-types}
\ee
Here, all the columns are equal to one of two linearly independent column vectors. Columns two and three are identical, so we can use column addition to eliminate one of them, leaving us with only two non-zero columns. As we will discuss shortly, this structure constitutes type 1. The rows, on the other hand, are equal to one of three different vectors. The last two rows are identical, so we can eliminate one of them with row addition, say the last one. Then we are left with the first three rows as the non-zero rows, but the sum of all three is zero. This allows us to eliminate one more row. This is clearly a different structure from the columns, and constitutes what we will shortly describe as type 2.
The type 1 and type 2 classifications are therefore properties of individual subsystems of a bipartite graph, not of the whole graph.

\subsubsection{Type 1 subsystem: two types of rows}
\label{type1-subsubsection}

In this type, all non-zero rows of the biadjacency matrix $G_{AB}$ can be divided into two sets of identical rows, such that the rows of one set are not equal to those of the other set. We can take the first row in set 1 and add it to all the other rows of the same set to turn them into zero rows. A similar operation can be performed for set 2. This implies that we have two distinct non-zero rows, and hence the matrix has rank 2. 

Let there be $n_1$ and $n_2$ rows in the two sets with
$n_A =n_1+n_2$. Physically, this means that $n_1$ of Alice's qubits have one set of identical ($CZ$) connections with Bob's qubits, and the remaining $n_2$ of her qubits have another identical set of connections. We will therefore call the choice of $n_1$ and $n_2$ a \emph{distribution}. Note that a distribution is independent of the details of the connections of the qubits in the two sets.

Now, since the density matrix of the full system is a mixture of several pure states, the reduced state for subsystem $A$ will have contributions from the reduced state associated with each of these full system pure states.
The reduced density matrix for Alice associated with a given pure state will be a product of two separate density operators each of the form arising in the rank 1 case, 
\begin{align}
\rho_{A, \vec m_A}
=&~ \frac{1}{2^{n_A}}\left[\prod_{j=2}^{n_1} (1+ (-1)^{m_{A, j-1}} 
X_{1} X_{j})\right]\nonumber \\
&\times\left[\prod_{k=2}^{n_2} (1+ (-1)^{m_{A, n_1+k-2}} 
X_{n_1+1} X_{n_1+k})\right]. \nonumber
\end{align}
The Von Neumann entropy of $\rho_{A,\vec m_A}$ is therefore the sum of the Von Neumann entropies of two rank 1 subsystems with $n_1$ and $n_2$ qubits. Recalling the result from Eq.~\eqref{rank1-entropy} for the rank 1 case, we obtain
\be
H(\rho_A) = f_1(n_1, P) + f_1(n_2, P).
\ee
The coherent information $\sI_A$ is then
\begin{align}
\sI_A &=\,f_1(n_1, P) + f_1(n_2, P) -n H_2(P) \nn \\
&=\,[f_1(n_1, P) -n_1 H_2(P)] \nonumber \\
& \quad\quad \mbox{} +  [f_1(n_2, P) -n_2H_2(P)] \,\,-n_B H_2(P)
\label{rank2-coherent-information-type1}
\end{align}
Thus the coherent information depends only on the distribution i.e. the choice of $n_1$ and $n_2$, and not the details of the connections between the two sets of Alice's qubits with those in Bob's subsystem.

\begin{figure}
\begin{subfigure}{.5\textwidth}
  \centering
  \includegraphics[width=\linewidth, trim={.5cm 0.25cm 0 0},clip]{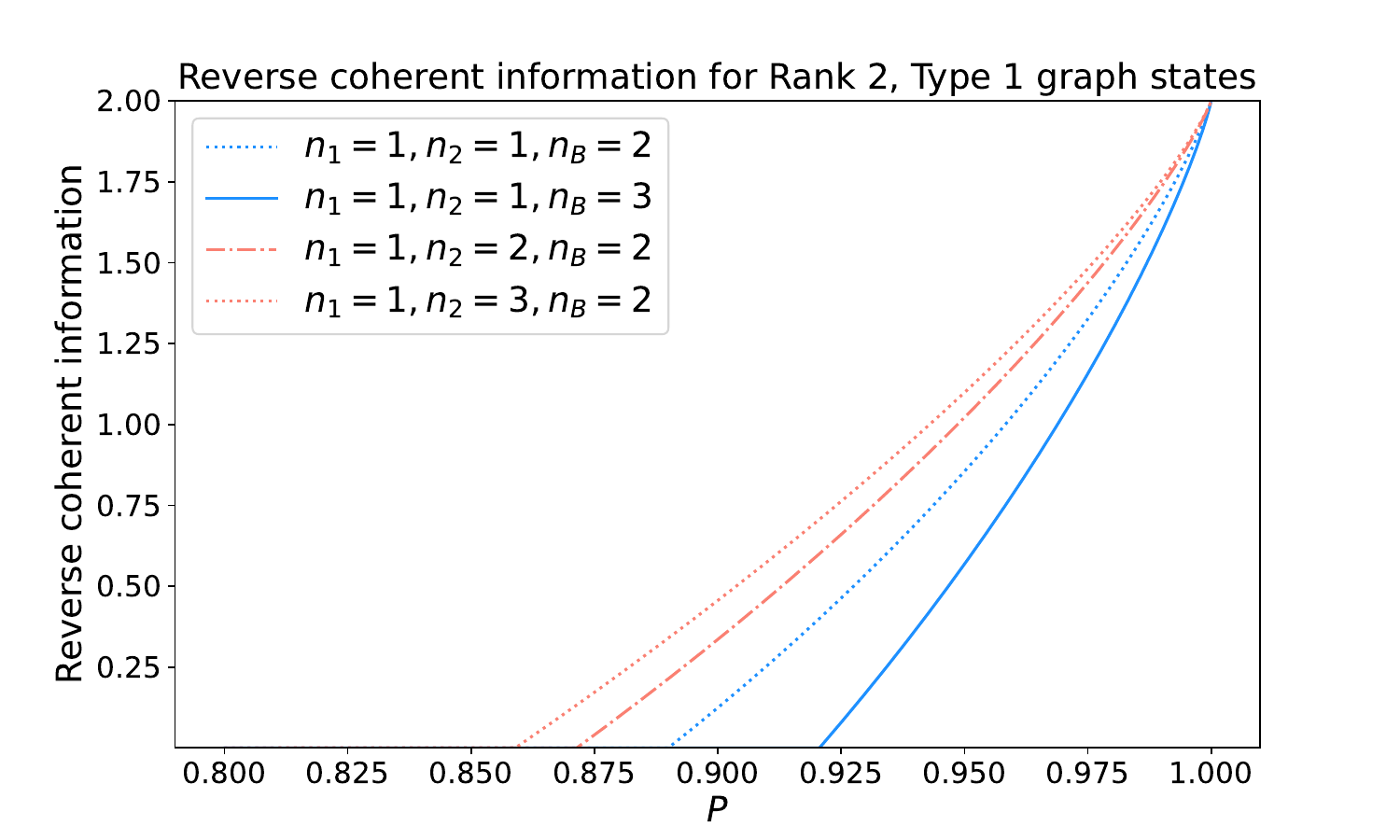}
\end{subfigure}

\begin{subfigure}{.5\textwidth}
  \centering
  \includegraphics[width=\linewidth, trim={.5cm 0.25cm 0 0},clip]{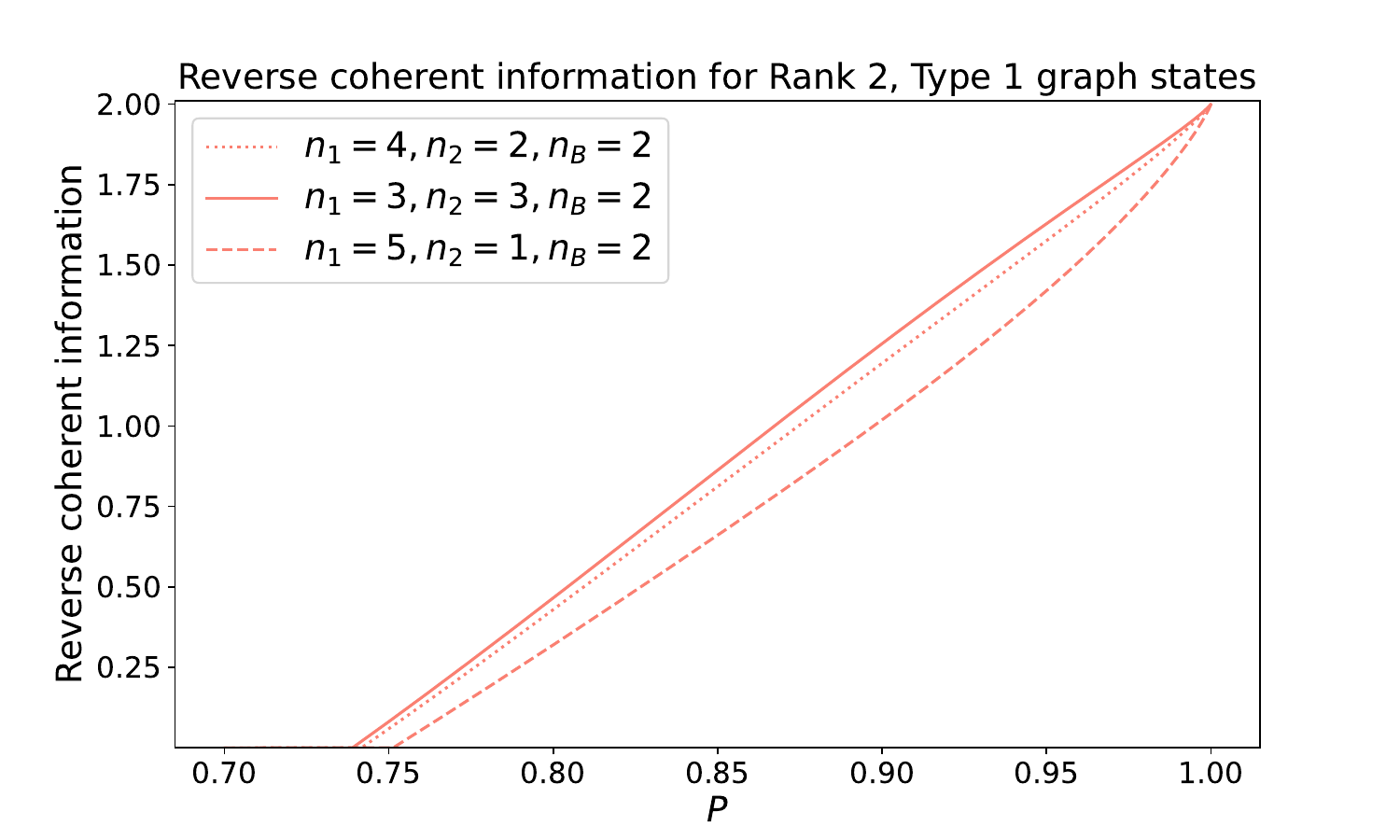}
\end{subfigure}
\caption{In these plots we see the impact of changing $n_1$, $n_2$ and $n_B$ on the coherent information $\sI_A$ for the rank 2, type 1 case given in Eq~\ref{rank2-coherent-information-type1}. For the first plot, we see the effect of increasing the number of Bob's qubits $n_B$ and Alice's qubits $n_1+n_2 = n_A$ on the coherent information $\sI_A$ for the rank 2, type 1 case. In the second plot, we find the optimal distribution of Alice's qubits. $P$ corresponds to the noise parameter.}
\label{fig:rank_2_type_1}
\end{figure}

See Fig~\ref{fig:rank_2_type_1} for details on the impact of the distribution of $n_1$, $n_2$, and $n_B$ on the coherent information $\sI_A$. 

It is easy to see the behavior of $\sI_A$ as $n_1$ and $n_2$ or $n_B$ and $n_2$ change. First, when $n_1 =n_2 = 1$,
we have $n_A=2$, and the reduced state of the subsystem is just the maximally mixed state on two qubits with entropy $2$. The coherent information $\sI_A$ in this case is thus simply $2-n H_2(P)$.
Beyond this, the behavior as we change $n_1$, $n_2$ and $n_B$ is as follows:
\begin{enumerate}
\item Given $n_1$ and $n_2$, $\sI_A$ is a decreasing function of $n_B$. 
\item Given $n_1$ and $n_B$, $\sI_A$ is an increasing function of $n_2$. Likewise, $\sI_A$ is an increasing function of $n_2$ for a given $n_2$ and $n_B$.
\item  Given $n_B$, $\sI_A$ is an increasing function of $n_1$ and $n_2$. Moreover, $\sI_A$ asymptotically approaches a constant for a given $P$.
\item For $n_B=2$, we obtain a purely positive coherent information for any non-maximal value of $P$ by taking $n_1$ and $n_2$ to be sufficiently large. This property directly follows from the fact that the coherent information for such a case can be expressed as the sum of the coherent informations of two star graphs.

\item Given $n_A$ and $n_B$, $n_1 = \lfloor n_A/2 \rfloor$ and $n_2 = \lceil n_A/2 \rceil$ maximizes $\sI_A$ over all values of $n_1$ and $n_2$ such that $n_1 + n_2 = n_A$. We call such a distribution of $n_A$, $n_B$ an \emph{equitable or optimal distribution}.
\end{enumerate}

The last property easily follows from the behavior of $[f_1(n_1, P) -n_1 H_2(P)]$ and $[f_1(n_2, P) -n_2 H_2(p)]$.
As we found during the rank 1 discussion, $[f_1(n_1, P) -n_1H_2(P)$ increases with $n_1$, but the rate of increase gradually slows down for larger $n_1$, and eventually $[f_1(n_1, P)-n_1H_2(P)]$ approaches a constant with respect to $n_1$.
Thus, if $n_1$ and $n_2$ are off by more than 1,
then decreasing $n_1$ by 1 and increasing $n_2$ by 1 should increase the sum $[f_1(n_1, P) -n_1H_2(P)] +[f_1(n_2, P) -n_2H_2(P)]$.
This continues until $|n_1-n_2| =1$ or $n_1=n_2$. Hence we conclude that an equitable 
distribution between sets 1 and 2 is optimal.


\subsubsection{Type 2 subsystem: three types of rows}
\label{type2-subsubsection}

In this type of subsystem, all row vectors of the biadjacency matrix $G_{AB}$ equal one of three distinct vectors $V_1$, $V_2$ and $V_3$. Secondly, these three vectors satisfy the property that their binary sum is the zero vector, i.e.~$V_1+V_2+V_3=0$.
Assuming that none of the qubits in the graph state are completely disconnected (i.e. no zero row or column in the biadjacency matrix), this property can only be satisfied if one and only one of the three vectors $V_1$, $V_2$ and $V_3$ comprises entirely of ones and the other two are $1$s complements of each other (i.e. flipping the $0$s to $1$s and $1$s to $0$s in one of these two vectors gives the other vector and vice versa).
Now, within each set of identical rows, we can take any row and add it to all the others eliminating all but one in the set. This way, we are left with just one non-zero row in each set (of originally identical rows). But since the sum of all three of them is zero, we can eliminate one of them by adding it to the other two. We are thus left with two linearly independent non-zero rows.

Assume that the number of rows belonging to each set is $n_1$, $n_2$ and $n_3$ with $n_A=n_1+n_2+n_3$. Thus Alice's qubits can be divided into three sets comprising $n_1$, $n_2$ and $n_3$ qubits. Each qubit within each such set has an identical connections with Bob's qubits specified by the three distinct row vectors whose sum is zero. In the same spirit as the type 1 case, we will call the choice of $n_1$, $n_2$ and $n_3$ a \emph{distribution}.

In Appendix \ref{sec:subsystem_entropy_calc} we present the calculations for the subsystem entropy. We find that it is given by
\begin{gather}
H(\rho_A) = f_2(n_1, n_2, n_3, P) \nonumber \\
\equiv 2-\sum_{n_{1+}, n_{2+}, n_{3+}=0}^{n_1-1, n_2-1, n_3-1} \binom{n_1-1}{n_{1+}}\binom{n_2-1}{n_{2+}} \binom{n_3-1}{n_{3+}}\nonumber\\
 \mbox{} \cdot (w_{\vec m_A, n_1, n_2, n_3, +, n_{1+}, n_{2+}, n_{3+}} \log\left(w_{\vec m_A, n_1, n_2, n_3, +, n_{1+}, n_{2+}, n_{3+}}\right)\nonumber \\
\mbox{} + w_{\vec m_A, n_1, n_2, n_3, -, n_{1+}, n_{2+}, n_{3+}}\log\left(w_{\vec m_A, n_1, n_2, n_3, -, n_{1+}, n_{2+}, n_{3+}} \right))\ \nonumber,
\end{gather}
where $w_{\vec m_A, n_1, n_2, n_3, +, n_{1+}, n_{2+}, n_{3+}}$ and $w_{\vec m_A, n_1, n_2, n_3, -, n_{1+}, n_{2+}, n_{3+}}$ are defined in Appendix \ref{sec:subsystem_entropy_calc}.

Therefore, the coherent information is given by
\begin{align}
\sI_A = & \;H(\rho_A)-H(\rho)\\
= & \;f_2(n_1, n_2, n_3, P) -n H_2(P)
\label{subsystem-entropy-type2-rank2} \ .
\end{align}
Again, like the coherent information for a type 1 subsystem, this depends only on the distribution (i.e. the choice of $n_1$, $n_2$ and $n_3$) and not the details of the exact connections of the qubits within each of the three sets.

\begin{figure}
\begin{subfigure}{.5\textwidth}
  \centering
  \includegraphics[width=\linewidth]{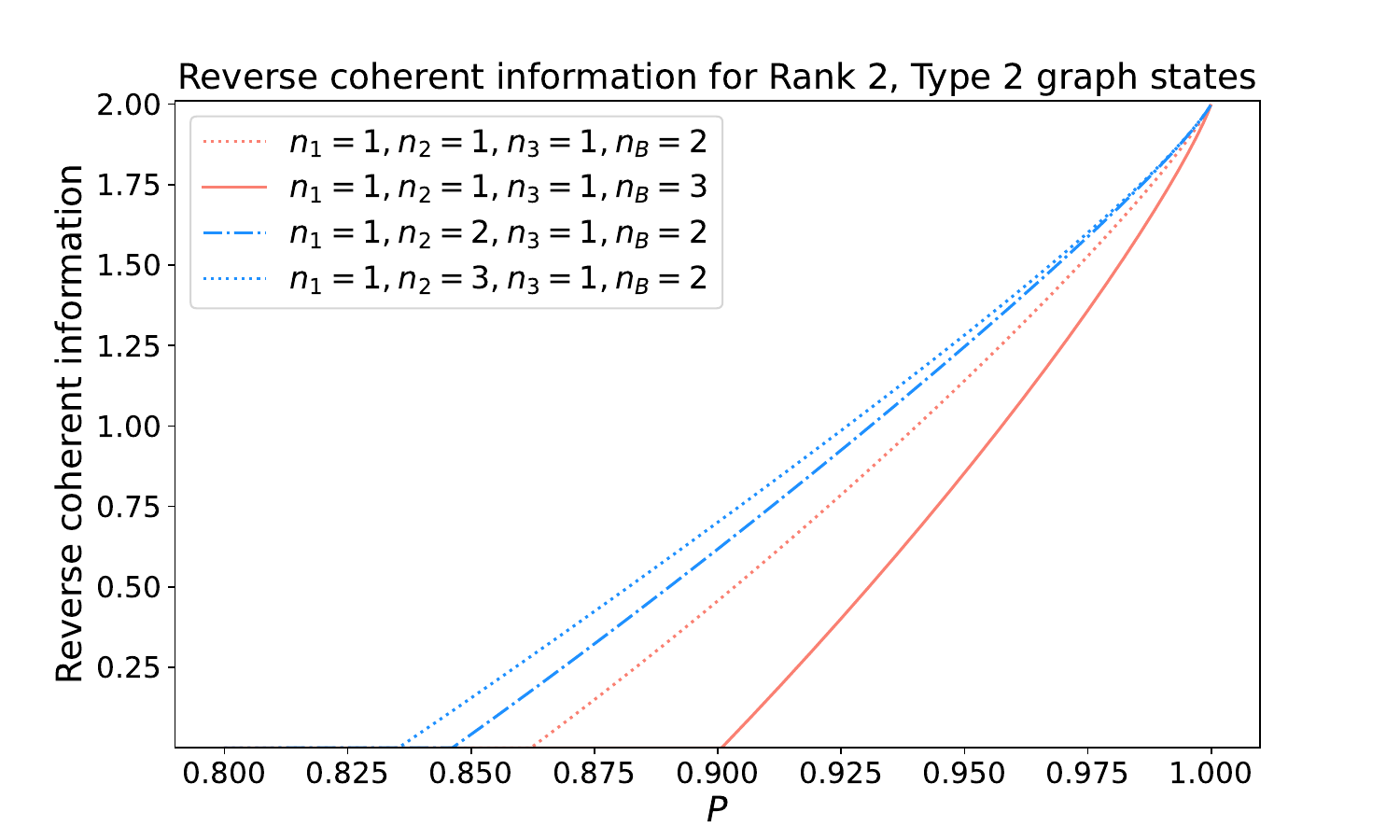}
\end{subfigure}

\begin{subfigure}{.5\textwidth}
  \centering
  \includegraphics[width=\linewidth]{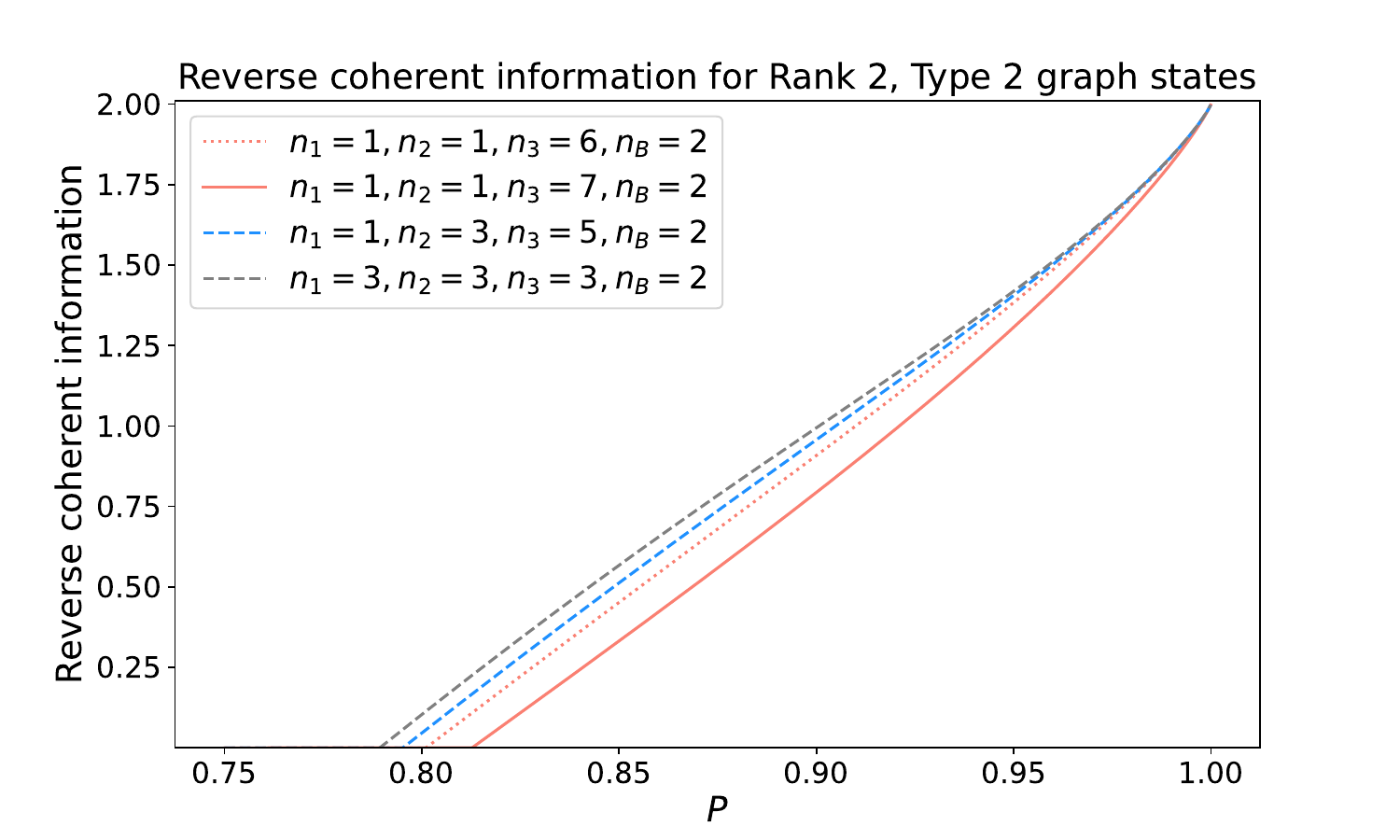}
\end{subfigure}
\caption{In these plots we see the impact of changing $n_1$, $n_2$ and $n_B$ on the coherent information for the rank 2, type 2 case given in Eq~\ref{subsystem-entropy-type2-rank2}. For the first plot, we see the effect of increasing the number of Bob's qubits $n_B$ and Alice's qubits $n_1+n_2+n_3 = n_A$ on the coherent information for the rank 2, type 2 case. In the second plot, we see the optimal distribution of Alice's qubits. $P$ corresponds to the noise parameter. }
\label{fig:rank_2_type_2}
\end{figure}

These results exhibit the following behavior:
\begin{enumerate}
\item Given $n_1$, $n_2$, and $n_3$, $\sI_A$ is a decreasing function of $n_B$. 
\item Given $n_1$, $n_3$, and $n_B$, $\sI_A$ is an increasing function of $n_2$. 
\item Keeping $n_B$ constant, and $n_A$ as the total number of Alice's qubits, if $n_A$ is a multiple of 3, $n_1=n_2=n_3 = n_A/3$, gives the highest coherent information $\sI_A$. 
If $n_A \mathrm{mod} 3 = 2$, $n_1 = (n_A-2)/3+1$, $n_2 = (n_A-2)/3+1$ and $n_3 = (n_A-2)/3$ gives the highest coherent information.  
If $n_A \mathrm{mod} 3 = 1$, $n_1 = (n_A-1)/3+1$, $n_2 = (n_A-1)/3$ and $n_3 = (n_A-1)/3$ gives the highest coherent information. We call such a distribution an \emph{equitable or optimal distribution} for type 2.
For details, see Fig.~\ref{fig:rank_2_type_2}.
\item If we fix $n_B=2$, and increase $n_A$ with an equitable distribution, we obtain purely positive coherent information for sufficiently large $n_A$ for any non-maximal noise parameter $P$. Thus the family of graphs with purely positive coherent information (for non-maximal noise) is larger than just star graphs or rank 2 type one graphs with $n_B=2$. More on this in section \ref{type1-and2-comparison-section} where we compare type 1 and type 2, and section \ref{higher-rank-generalization-section} where we discuss the generalizations to higher ranks.
\item The above robustness (i.e. purely positive coherent information for any non-maximal noise for sufficiently large $n_A$) does not however hold in general if we consider inequitable distributions instead of equitable ones.
\end{enumerate}

\subsubsection{Rank 2 graphs: considering both systems together}

Having described types 1 and 2 for the individual subsystems, we can now consider them together to categorize graphs with rank 2 biadjacency matrices. Since either of the two subsystems can be of type one or two, there are four possible forms of rank 2 graphs.

\subsubsection*{When both subsystems are type 1}

Let us say Alice's qubits are divided into two sets $A_1$ and $A_2$ with $n_{1_A}$ and $n_{2_A}$ qubits, respectively. Likewise, assume that Bob's qubits are divided into two sets $B_1$ and $B_2$ having $n_{1_B}$ and $n_{2_B}$ qubits, respectively. We can have a rank 2 graph with type 1 for both subsystems when the biadjacency matrix (after possible relabling of the qubits) is given by one of the following:
\be
\begin{pmatrix}
V(n_{1_A}, n_{1_B}) & 0 \\
0 & V(n_{2_A}, n_{2_B})
\end{pmatrix},
\ee
\be
\begin{pmatrix}
0 & V(n_{1_A}, n_{2_B}) \\
V(n_{2_A}, n_{1_B}) & 0
\end{pmatrix},
\ee
\be
\begin{pmatrix}
V(n_{1_A}, n_{1_B}) & 0 \\
V(n_{2_A}, n_{1_B}) & V(n_{2_A}, n_{2_B})
\end{pmatrix},
\ee
\be
\begin{pmatrix}
V(n_{1_A}, n_{1_B}) & V(n_{1_A}, n_{2_B} \\
0 & V(n_{2_A}, n_{2_B})
\end{pmatrix},
\ee
\be
\begin{pmatrix}
0 & V(n_{1_A}, n_{2_B}) \\
V(n_{2_A}, n_{1_B}) & V(n_{2_A}, n_{2_B})
\end{pmatrix},
\ee
\be
\begin{pmatrix}
V(n_{1_A}, n_{1_B}) & V(n_{1_A}, n_{2_B}) \\
V(n_{2_A}, n_{1_B}) & 0
\end{pmatrix},
\ee
where $V(i, j)$ is an $i\times j$ matrix comprising entirely of $1$s, and $0$ denotes a matrix (of an appropriate size) comprising entirely of zeros.
The top left block specifies the connections between qubits in $A_1$ and $B_1$, and the top right gives the edges between qubits in $A_1$ with $B_2$. Likewise, the bottom left and bottom right blocks give the edges between $A_2$ and $B_1$, and $A_2$ and $B_2$ qubits, respectively.

In each of these six graphs, we have the type 1 distribution for Alice's subsystem given by $n_{1_A}$ and $n_{2_A}$, and the type 1 distribution for Bob specified by $n_{1_B}$ and $n_{2_B}$. In fact these six graphs are the only ones having these particular distributions for the two subsystems. With the same distributions, these graphs yield the same pair of $\sI_A$ and $\sI_B$ values for the two coherent informations.

\subsubsection*{When one system is type 1, the other type 2}

Here we focus on the case where Alice's subsystem is type 2 and Bob has type 1. The opposite scenario will have a similar structure. Let us say that Alice's qubits are divided into three sets $A_1$, $A_2$ and $A_3$ with $n_{1_A}$, $n_{2_A}$ and $n_{3_A}$ qubits, respectively. Likewise, assume that Bob's qubits are divided into two sets $B_1$ and $B_2$ with $n_{1_B}$ and $n_{2_B}$ qubits, respectively.
We have a rank 2 graph with type 2 for Alice with the distribution given by $n_{1_A}$, $n_{2_A}$ and $n_{3_A}$,
and type 1 for Bob with the distribution given by $n_{1_B}$ and $n_{2_B}$, for the following two biadjacency matrices:
\be
\begin{pmatrix}
0 & V(n_{1_A}, n_{2_B}) \\
V(n_{2_A}, n_{1_B}) & V(n_{2_A}, n_{2_B}) \\
V(n_{3_A}, n_{1_B}) & 0
\end{pmatrix}
\ee
 and
\be
\begin{pmatrix}
V(n_{1_A}, n_{1_B}) & 0 \\
V(n_{2_A}, n_{1_B}) & V(n_{2_A}, n_{2_B}) \\
0 & V(n_{3_A}, n_{2_B})
\end{pmatrix},
\ee
where the top, bottom and middle rows of the block form describe the edges of the qubits in $A_1$, $A_2$, and $A_3$, respectively, and the columns pertain to $B_1$ and $B_2$.
Up to possible relabeling of qubits, these two graphs are the only ones with the above-mentioned distributions for the two subsystems,
and yield the same pair of values for the two coherent informations $\sI_A$ and $\sI_B$ corresponding to these distributions.

\subsubsection*{When both subsystems are type 2}

Assume that Alice's qubits are divided into three sets $A_1$, $A_2$ and $A_3$ with $n_{1_A}$, $n_{2_A}$ and $n_{3_A}$ qubits, respectively. Likewise, let us say that Bob's qubits are divided into three sets $B_1$, $B_2$ and $B_3$ with $n_{1_B}$, $n_{2_B}$ and $n_{3_B}$ qubits, respectively.
We have a rank 2 graph with type 2 for both subsystems for the following two biadjacency matrices:
\be
\begin{pmatrix}
0 & V(n_{1_A}, n_{2_B}) & V(n_{1_A}, n_{3_B}) \\
V(n_{2_A}, n_{1_B}) & V(n_{2_A}, n_{2_B}) & V(n_{2_A}, n_{2_B}) \\
V(n_{3_A}, n_{1_B}) & V(n_{3_A}, n_{2_B}) & 0
\end{pmatrix}
\ee
 and
\be
\begin{pmatrix}
V(n_{1_A}, n_{1_B}) & V(n_{1_A}, n_{2_B}) & 0 \\
V(n_{2_A}, n_{1_B}) & V(n_{2_A}, n_{2_B}) & V(n_{2_A}, n_{3_B}) \\
0 & V(n_{3_A}, n_{2_B}) & V(n_{3_A}, n_{3_B})
\end{pmatrix},
\ee
Thus both these graphs have a type 2 distribution for Alice with the numbers $n_{1_A}$, $n_{2_A}$, and $n_{3_A}$, and a type two distribution for Bob's subsystem with $n_{1_B}$, $n_{2_B}$ and $n_{3_B}$ for the three subsets of Bob's qubits. These two graphs therefore give the same pair of coherent informations $\sI_A$ and $\sI_B$ values for both subsystems. And (up to possible relabling of qubits,) these are the only two graphs giving these particular distributions.
 
\subsubsection*{Properties}

The properties of the above three cases (involving both subsystems) for rank 2 graphs depend on the distribution parameters for both subsystems and their types. These include the properties for types 1 and 2 subsystems described in the previous two sections as well as the comparisons in the next one.
We should also remind the reader that the larger of the two coherent informations $\sI_A$ and $\sI_B$ gives a tighter bound on the bipartite entanglement. For all the many examples we have calculated, we have found a clear pattern that the subsystem with the larger number of qubits yields a larger coherent information, regardless of the types of the subsystems. We therefore believe this is a general property, but do not have a rigorous proof. We offer this as a conjecture in section \ref{higher-CI-for-more-qubit-subsystem}.

\section{Comparison between graph states}
\label{sec:results}
With the analytical results at our disposal, we can study the robustness of different rank 1 and rank 2 graph states. We first compare the two types of rank 2 graph states, and then compare rank 1 and rank 2 graph states. We close with a brief discussion on how to potentially generalize our approach to higher ranks.

\subsection{Comparing type 1 and type 2 subsystems}
\label{type1-and2-comparison-section}

\begin{figure}
    \includegraphics[width = .5\textwidth,trim={1.0cm 0.5cm 0 0},clip]{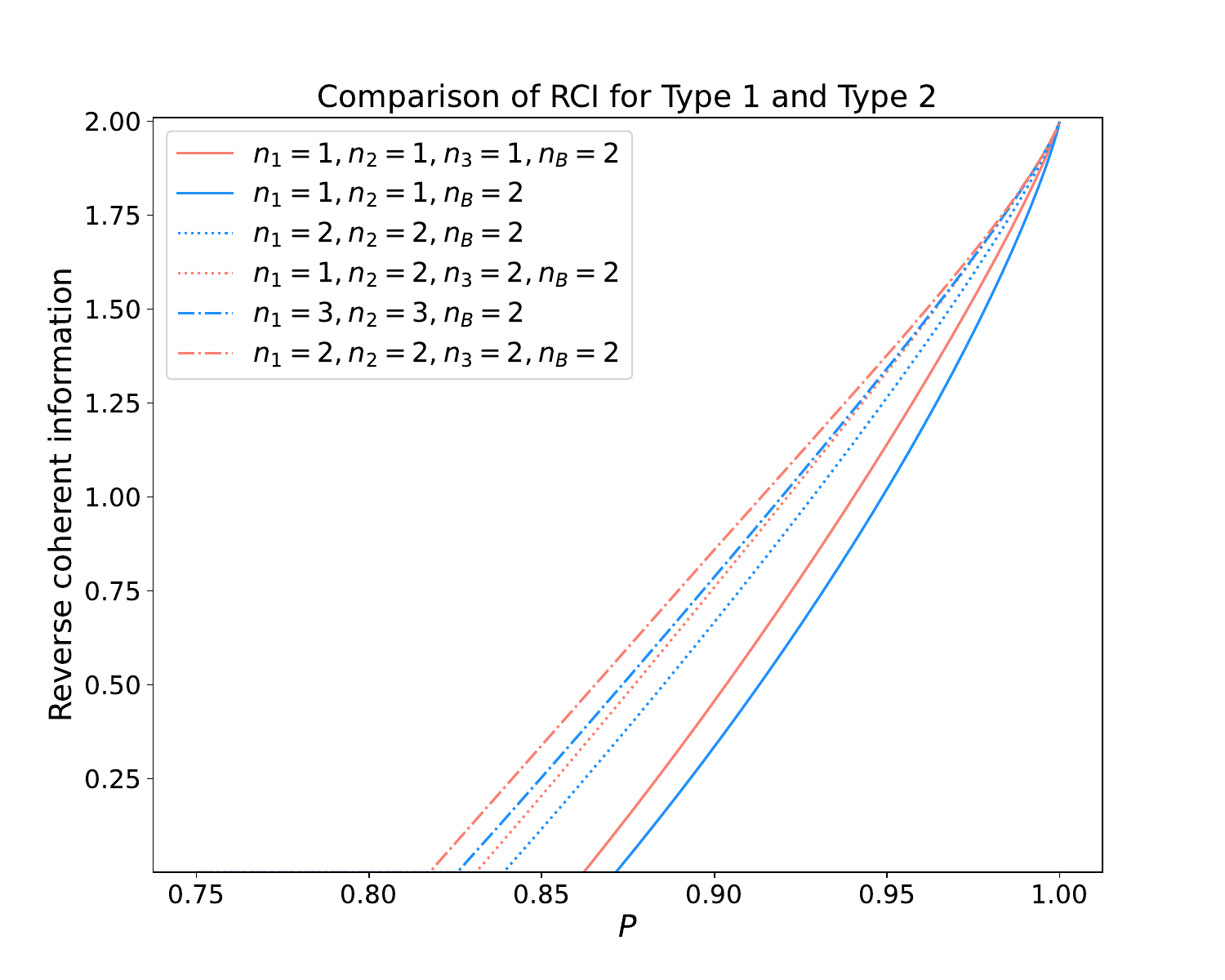}
    \caption{In this plot, we compare various optimal Type 1 and Type 2 configurations as given in Sec~\ref{type1-and2-comparison-section}. Type 1 is denoted by the blue lines, while Type 2 is denoted by the red lines. We also plot a non-optimal configuration for Type 2 and see that it performs worse than the optimal Type 1. $n_B$ refers to the number of qubits with Bob. Type 1 is characterized by $n_1, n_2$, where $n_1+n_2 = n_A$. Type 2 is characterized by $n_1,n_2$ and $n_3$, where $n_1+n_2+n_3 = n_A$, where $n_A$ is the number of qubits with Alice. $n_B$ represents the number of qubits held by Bob and $P$ corresponds to the noise parameter.} 
    \label{fig:comparison}
\end{figure}

Take $n_A$ and $n_B$ to be constant. We find that the optimal distribution  for a type 2 subsystem always does better than the optimal distribution for a type 1 subsystem as seen in Figure~\ref{fig:comparison}.
This also explains the property that for fixed $n_B=2$, the equitable type 2 distribution belongs to the family of graphs with positive coherent information for any non-maximal noise when we take $n_A$ to be sufficiently large.
We have already shown that we obtain such robustness to noise for fixed $n_B=2$ when Alice's subsystem is of type 1. Therefore, the fact that the optimal distribution for type 2 always yields higher coherent information than the optimal distribution for type 1, means that the former will also have a purely positive coherent information for any non-maximal noise for a sufficiently large $n_B$.

However, we can find some distributions for type 1 that do better than some distributions of type 2 for equal values of $n_A$. For example, the type 2 case with $n_1 =1$, $n_2= 1$, $n_3 = 4$, $n_B =2$ performs worse than the type 1 case of $n_1 =3$, $n_2 =3$ and $n_B =2$.

\subsection{Comparing rank 1 and rank 2}
\label{rank1-and-2-comparison-section}

\begin{figure}
\includegraphics[scale = .35]{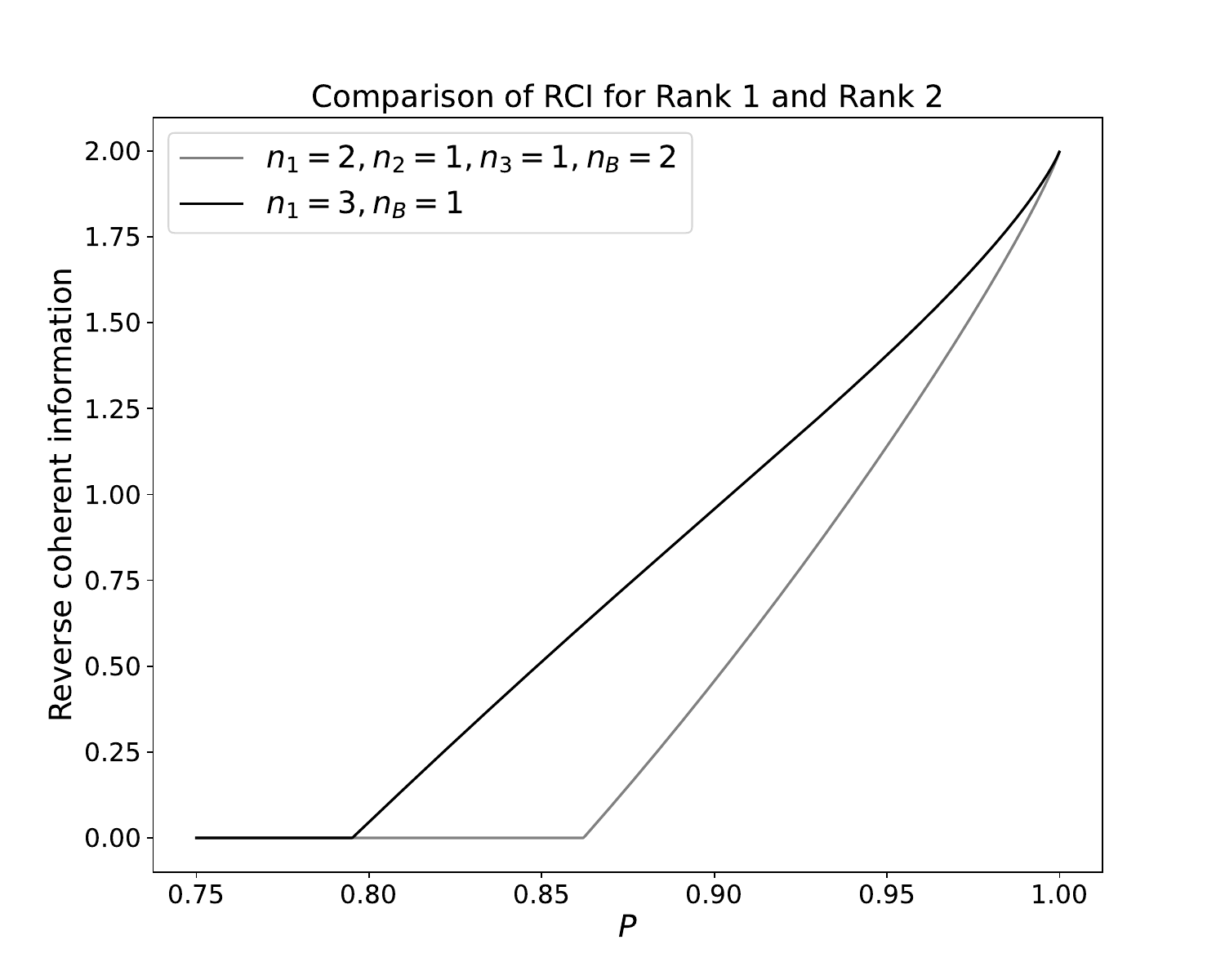}
    \caption{In this plot, we compare the Coherent Information for rank 1 and rank 2 cases. Here, $n_B$ represents the number of qubits held by Bob and $P$ corresponds to the noise parameter.}
    \label{fig:comparison_ranks}
\end{figure}

In this section, we compare the Coherent Information for the rank 1 case with that of rank 2. We start by fixing $n_B = 1$ for rank 1 and $n_B = 2$ for rank 2. We fix $n_A = 3$. We then compare the optimal distribution for rank 2 with the rank 1 case. We observe that the rank 2 case performs better than rank 1 for larger $P$ and vice versa. See Fig.~\ref{fig:comparison_ranks} for details. This behavior for large $P$ is to be expected since $P\to1$ is the noiseless limit. It is however interesting that we can obtain higher coherent information for rank 1 compared to rank 2 as the noise increases. 

\subsection{The generalization to higher ranks}
\label{higher-rank-generalization-section}

The generalization to higher ranks is more complicated. For example, for rank 3, we have three types of subsystems based on extensions of the two rank 2 types. But then we can also have another more complicated type as well. 

Let us first begin with the first three types. Recall that for rank 2, we have a type 1 subsystemwhere all the rows of the biadjacency matrix are equal to one of two linearly independent vectors. In each set of identical rows, we can eliminate all but one row through row operations, leaving a total of two non-zero rows in the row echelon form. In the same spirit, we have rank 3 if all the rows are equal to one of three linearly independent vectors. Within each set, we can eliminate all rows except one, leaving a total of three linearly independent rows, and thus giving rank 3. We can call this rank 3 type 1. Each set of identical rows would in isolation give a rank 1 graph if the qubits corresponding to the remaining rows were to be removed. The (subsystem) entropy of a rank 3 type 1 subsystem will therefore be equal to the sum of the three rank 1 pieces i.e. $f_1(n_1, P) + f_1(n_2, P) + f_1(n_3, P)$, where $n_1$, $n_2$ and $n_3$ are the number of rows in the three linearly independent sets.

A second type (for rank 3) can be constructed out of rank 1 and rank 2 type 2 pieces. Say all the rows are equal to one of four distinct row vectors $V_1$, $V_2$, $V_3$ and $V_4$. If one of them, say $V_1$, is linearly independent from the other three and $V_2+V_3+V_4=0$, then we get rank 3. The set of rows equal to $V_1$ then constitute a rank 1 piece (as we can eliminate all but one of them through row additions), and the remaining rows constitute a rank 2 piece of type 2. That is, if we were to remove all the qubits in the graph corresponding to the rows equal to $V_1$, then we would be left with a rank 2 type 2 subsystem for Alice. This way, the overall entropy of such a subsystem is the sum 
$f_1(n_1, P) + f_2(n_2, n_3, n_4, P)$,
assuming that the rank 1 subset has $n_1$ rows, and the rank 2 type 2 set has $n_2$, $n_3$ and $n_4$ rows in each of its three sets.
Naturally, such rank 3 type 2 cases inherit the properties of rank 1 and rank 2 type 2 entropies.

Then there will be a type 3 for obtaining rank 3, in which we will obtain a new expression for the subsystem entropy instead of comprising a sum of rank 1 and rank 2 submatrices. This will be similar to rank 2 type 2. That is, now 
all rows of the biadjacency matrix are equal to one of four vectors which have the property that adding all four of them will give a zero row, but if we take any three of them, then they are linearly independent from each other. 
We can use Gaussian elimination to eliminate all but one row within each set of identical rows. After that, we can eliminate one of the surviving four rows by adding to it the other three. The entropy $f_3(n_1, n_2, n_3, n_4, P)$ of this can be calculated by going through all the combinatorics similar to the rank 2 type 2 case we have explicitly worked out.

While these are the three cases similar to the two rank 2 types, there is also a fourth possibility. Let us say we have five different row vectors $v_1\ldots v_5$ with the properties that the binary sum $v_1+v_2+v_3$ gives a zero row, and similarly,
$v_1+v_4+v_5$ is also a zero vector. This way, both binary sums eliminate one vector each, and we are left with three linearly independent vectors giving rank 3. A biadjacency matrix whose row vectors are all equal to five vectors satisfying the above properties will thus have rank 3. A specific example of this is the $5\times 4$ biadjacency matrix
\be
\begin{pmatrix}
1 & 1 & 1 & 0 \\
1 & 1& 1 & 1 \\
0 & 0 & 0 & 1 \\
1 & 0 & 0 & 1 \\
0 & 1 & 1 & 1 \\
\end{pmatrix}
\ee
Here the sum of rows 1, 2 and 3 is zero, and the sum of rows 1, 4 and 5 is also zero, and we have type 4 rank 3 in terms of the rows. As for the columns, the second and third columns are identical, and these are linearly independent from columns 1 and 4. Thus we have type 1 in terms of the columns; we just add columns 2 to 3 and replace one of them by the sum which is a zero column, and we are left with three linearly independent columns.

In the same way, we can consider even higher ranks, but these become more and more complicated. A detailed exploration of ranks larger than 2 is beyond the scope of this work and we therefore do not discuss these any further. However, we point out that for large ranks close to the number of qubits $n_A$ or $n_B$ in a subsystem, it can be easier to calculate the subsystem entropy by working out the combinatorics in terms of $K_A$ or $K_B$. We show how to calculate the subsystem entropy for the cases $K_A=1$ and $K_A=2$ in Appendices
\ref{K_A=1-section} and \ref{K_A=2-section}.

Before closing this section, we would also like to address what the above discussion means for higher rank members of the family of robust graphs having purely positive coherent information for any non-maximal noise parameter $P$.
Recall that we found that star graphs with $n_B=1$ belong to this family for sufficiently large $n_A$. 
This also holds for the 
for the rank 2 type 1 case for Alice's subsystem with fixed $n_B=2$, since $\sI_A$ can then be expressed as the sum of two star graph coherent informations.
Moreover, we also found that the above-mentioned robustness property also holds if we have an equitable distribution for Alice's rank 2 type 2 subsystem (with $n_B=2$), but not for inequitable distributions.
Based on this pattern, it is clear that certain higher rank graphs where $n_B$ equals the rank will also belong to the family of graphs with this robustness to noise for large enough $n_A$.
Any graph whose rows of the biadjacency matrix can be separated into a combination of rank 1 and/or equitably distributed rank 2 type 2 submatrices will inherit the robustness property. For example, the type 1 and type 2 cases for rank 3 described above will belong in this category. As for other more complicated types, such as types 3 and 4 for rank 3 described above, we are not in a position to make any claim without calculating the coherent information and looking for patterns. However, based on the behavior of the rank 2 type 2 cases, we offer a conjecture that equitable distributions will perhaps satisfy the robustness property whereas inequitable distributions will not. Further investigation of this issue is however beyond the scope of this paper.

\subsection{The subsystem with more qubits has higher coherent information?}
\label{higher-CI-for-more-qubit-subsystem}

In all of the examples that we have considered, we have observed that the subsystem  with the most qubits has the largest coherent information regardless of the types of the two subsystems and their distributions. Therefore, we conjecture that this is generally true but are unable to make a definitive statement. 
We can however prove this for two special cases:
\begin{enumerate}
 \item We have already shown in Section \ref{rank1-section} that for any fully connected i.e. rank 1 graph, if $n_A> n_B$, then $\sI_A >\sI_B$. 
\item We can also prove this for the case where the number of qubits in one of the two subsystems, say $A$ equals the rank, so that $K_A=0$. Then $\rho_A$ is the maximally mixed state with entropy $n_A = n_B-K_B$. But the entropy for subsystem $B$ will be
\be 
n_B-K_B -\sum_{\vec m_B} w_{\vec m_B} \log_2(w_{\vec m_B}) \ . 
\ee
Subsystem $B$ will thus have a higher entropy due to the additional $-\sum_{\vec m_B} w_{\vec m_B} \log_2(w_{\vec m_B})$ contribution. 
\end{enumerate}

\section{Conclusion}
\label{sec:conclusion}
In this work we have studied the effects of noisy state preparation and dephasing on graph states. In particular, we were concerned with understanding the entanglement across a given bipartition. We described a general algorithm to calculate the coherent information, which---as a lower bound on the distillable entanglement---serves as a proxy for the entanglement in the states. While the algorithm is not efficient, we were still able to probe the behaviour of graph states that are expected to be implemented in the near-term future.

For certain families of graph states the coherent information can be calculated exactly. We found a class of graphs for which it is possible to tolerate any amount of (non-maximal) noise by introducing more qubits. This shows the practical relevance of our results; noise can be dealt with by adding an overhead, similar to error correction and distillation.

There still remain pressing open questions, however. Are there heuristics for finding graph states whose entanglement across a bipartition is robust against noise? What happens when depolarizing noise is applied \emph{after} the application of the $\CZ$ gates? What about non-Clifford noise that may emerge in realistic preparation of graph states among matter quantum memories~\cite{Dhara2023}? We provide a partial answer to these two questions in~\cite{goodenough2024bipartite}, showing that, after stabilizer twirling, the coherent information for any noise model can be interpreted in terms of syndrome entropies of certain stabilizer codes. 

\FloatBarrier
\section{Acknowledgements}
This work was funded by the Army Research Office (ARO) MURI Program Project on Quantum Network Science, ``Theory and Engineering of Large-Scale Distributed Entanglement", awarded under grant number W911NF2110325. AS and SG thank Liang Jiang, Yat Wong and Ashlesha Patil for helpful discussions.

\FloatBarrier
\newpage
\FloatBarrier
\appendix
\newpage
\section{An algorithm for general $K_A$}
\label{sec:algorithm}

Here we show how to calculate the Von Neumann entropy for a reduced subsystem arising from our noisy graph state. First, we construct the matrices $G_{\rm ext, A}$ and $G_{\rm ext, B}$ defined in (\ref{extended-biadjacency-matrices}), which we reproduce here
\bea
G_{\rm ext, A} &\equiv \begin{pmatrix}
I_A & G_{AB} 
\end{pmatrix} \nn\ , \\
G_{\rm ext, B} &\equiv \begin{pmatrix}
G_{AB} ^T & I_B
\end{pmatrix} \ ,
\eea
where $I_A$ and $I_B$ are $n_A\times n_A$ and $n_B\times n_B$ identity matrices, respectively, and $G_{AB}$ is the biadjacency matrix describing the edges across the bipartition in the graph.

We carry out a sequence of row operations on $G_{\rm ext, A}$ to reduce the $G_{AB}$ block to its row echelon form. Once this row reduction procedure is done, we identify the lower left $K_A \times n_A$ block of the final form of $G_{\rm ext, A}$ as the matrix $\sJ_A$ defined in Section \ref{useful-notation-section}.
Likewise, we apply row operations on $G_{\rm ext, B}$ to reduce the $G_{AB} ^T$ block to its row echelon form, and identify the lower right $K_B \times n_B$ block of the final form of $G_{\rm ext, B}$ as the matrix $\sJ_B$.

In what follows in the rest of this section, we mainly focus on calculating the Von Neumann entropy of Alice's subsystem with the understanding that the same procedure will give the entropy for Bob's reduced  system.

To speed up the algorithm, we can trim $\sJ_A$ by removing any zero columns.
Let us say we have a total of $\nu_A$ non-zero columns in $\sJ_A$.
and denote the trimmed matrix as $\sJ^\prime _A$.
(We can also work without this trimming, but that will involve going through the configurations of the qubits corresponding to the zero columns. Their total probability will just sum to one and the eventual result will have no dependence on them. Therefore dropping them speeds up the calculation.)
 
Define a vector $w_A$ with $2^{K_A}$ entries indexed by $l =0\ldots 2^{K_A} -1$. Each entry will store the probability associated with a separate bracket configuration given by $\vec m_A$ as we will show in the coming steps. At this stage, we initialize all the entries of $w_A$ as zero.

Define the $K_A$ component binary vector $\vec m_A$ for the different bracket configurations as described in Section \ref{useful-notation-section}.

Likewise, define a binary vector $\vec a$ with $\nu_A$ entries labeled with an index $i=0\ldots \nu_A-1$.
This will give the initial configuration of the relevant qubits with $a_i=0$ for the corresponding qubit being initialized in $|0\ra$ and $a_i=1$ for $|-\ra$.

To go through the various qubit configurations, run a loop over an integer $d$ going from $0$ to $2^{\nu_A}-1$. For each $d$ in this loop, 
\begin{enumerate}
\item Evaluate the binary representation of the integer $d$ and store it in the binary vector $\vec a$. The entries of $\vec a$ give the initial qubit configuration associated with integer $d$, with $0$ ($1$) representing $|+\ra$ ($|-\ra$) for each qubit $i$.
\item Calculate the probability $P_d$ of the qubit configuration associated with $d$. For this, first obtain the number of $1$ entries in $a_i$ as
\be
\nu_- = \sum_{i=0}^{\nu_A-1} a_i
\ee
Then the probability of the configuration $d$ is
\be
P_d = P^{\nu_A -\nu_-} (1-P)^{\nu_-}
\ee
\item From qubit configuration $\vec a$, obtain all the entries of $\vec m_A$ for the bracket configuration using
\be
m_{a, k} 
= \sum_{j=0}^{\nu_A -1}
a_j \sJ^\prime _{A, kj}
\, \textrm{ mod }\, 2 
\ee
\item Reading $\vec m_A$ as a binary string, calculate the integer for which it is the binary representation. Asign this value to the integer $l$.
\item Return to the vector $w$ defined earlier. Add to the existing value of the $l$'th entry of $w$ the probability $P_d$ of the qubit configuration $d$. That is,
\be
w_l = w_l + P_d
\ee
This way, as we run the loop over $d$ to go through all the qubit configurations, we iteratively add its probability contribution to $w_l$, which stores the probabilities for the different bracket configurations associated with $\vec m_A$.  
\end{enumerate}

When we are done with the loop over $d$, we have all the probabilities $w_l$ associated with the different bracket configurations. We then simply calculate the Von Neumann entropy from these as
\be
H(\rho_A) = n_A-K_A \,-\, \sum_{l=0}^{2^{K_A}-1} w_l \log_2(w_l)
\ee
This can be calculated iteratively by using a loop.

\section{Connections between star graphs and repetition code}
\label{star-graph-repetition-code-appendix}

In this section, we explore the relation between the repetition code and the star graph state -- a rank 1 graph state. Let us consider a two-qubit graph state. If we consider that both qubits have undergone the initialization error, then the resultant state is a Bell diagonal state. The coherent information is given by $1-2H(p)$, where $p$ is the depolarizing noise. 

We seek to modify the graph state to increase the coherent information. We start with a simple graph with $n$ qubits on Bob's side named $B_1, B_2, B_3 \cdots B_n$ and one qubit on Alice's side $A_1$. We have edges between $A_1 B_1$, $A_1 B_2$, $A_1B_3$ and so on.

For now, consider that only the qubits on Bob's side has the initialization error and call this state $\rho^n_{A_1B_1B_2\cdots B_n}$. We give a lower bound on the coherent information of $\rho^n_{A_1B_1B_2\cdots B_n}$ that tends to one as $n$ increases. This implies that even in the presence of noise, by increasing the number of qubits on Bob's side, the distillable entanglement of the state tends to one. The noise resilience of the state can be explained in terms of the repetition code.

Let $n=3$.
We attach two ancilla qubits $P_1$ and $P_2$, perform a Hadamard gate on $B_1B_2B_3$ and CNOT gate from $B_1$ to $P_1$, CNOT from $B_2$ to $P_1$, CNOT from $B_2$ to $P_2$ and CNOT from $B_3$ to $P_2$. We then perform a CNOT from $P_1$ to $P_2$. This implements a decoder for the repetition code. By performing error correction on $B_1$, $B_2$, and $B_3$ according to the syndromes we obtain and perform a $Z$ measurement on the qubits to remove the qubits $B_2$ and $B_3$. This gives us a Bell diagonal state with the coefficients:

\begin{align}
    p^3+ 3p^2(1-p) &: \ket{0+}+\ket{1-}, \\
    3p(1-p)^2 +(1-p)^3 &: \ket{0-}+\ket{1+}, 
\end{align}
from which we can calculate the coherent information of the state. 

We thus see that using the method given above, we can correct for at most $\lfloor \frac{n}{2}\rfloor$ errors for odd $n$. We can then easily write the expression for the resultant Bell diagonal state:

\begin{align}
    \sum_{k=0}^{\lfloor \frac{n}{2}\rfloor}\  ^nC_k p^{n-k}(1-p)^k&: \ket{0+}+\ket{1-},\\
    1-\sum_{k=0}^{\lfloor \frac{n}{2}\rfloor}\  ^nC_k p^{n-k}(1-p)^k &: \ket{0-}+\ket{0+}. 
\end{align}

\begin{figure}
\begin{subfigure}{.5\textwidth}
  \includegraphics[width=1\linewidth]{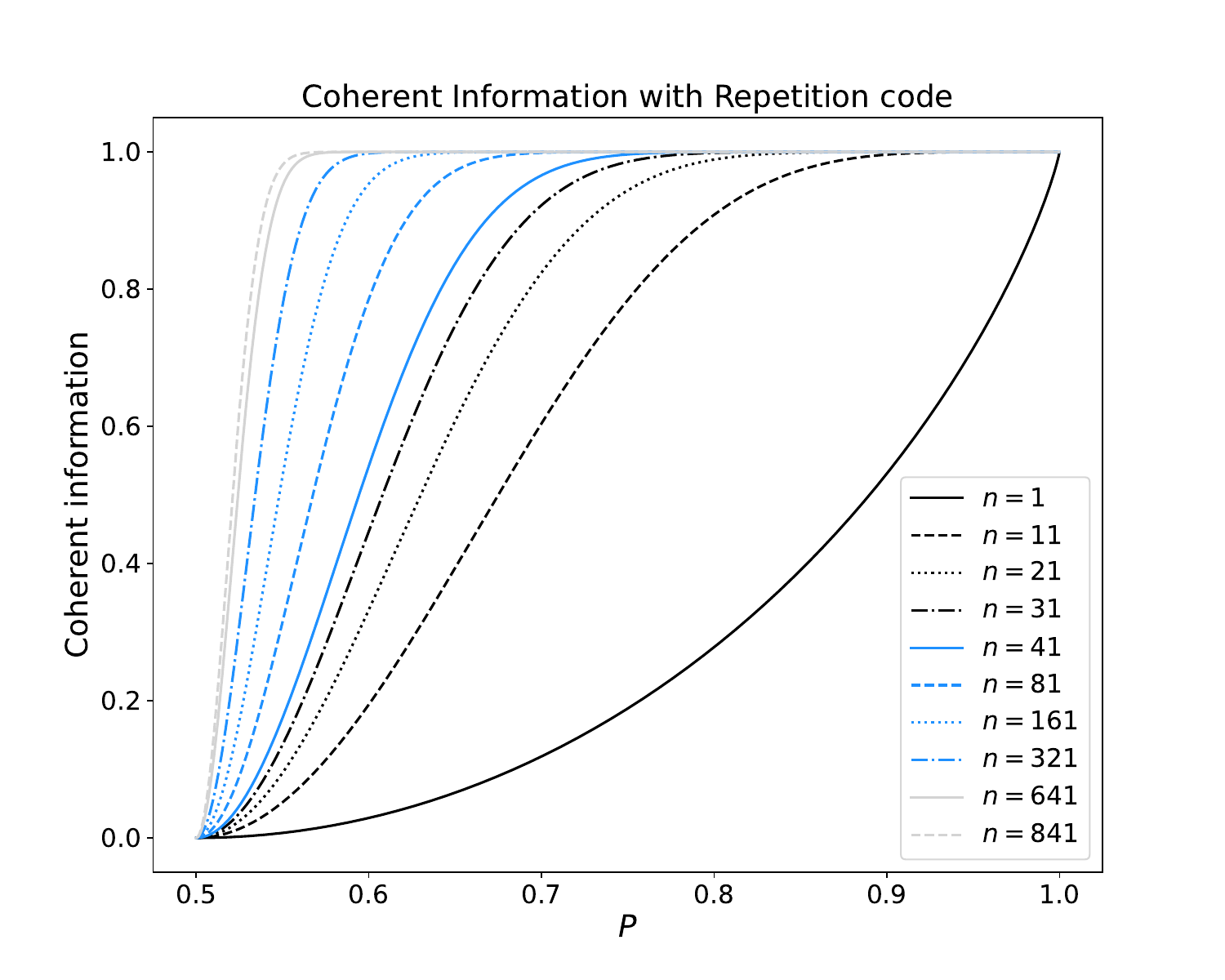}
  \caption{This plot gives the coherent information of state where one qubit is held with Alice and $n$ qubits are with Bob. The initialization errors are only on Bob's qubits.}
  \label{fig:rep_code_1}
\end{subfigure}

\begin{subfigure}{.5\textwidth}
  \centering
  \includegraphics[width=1\linewidth]{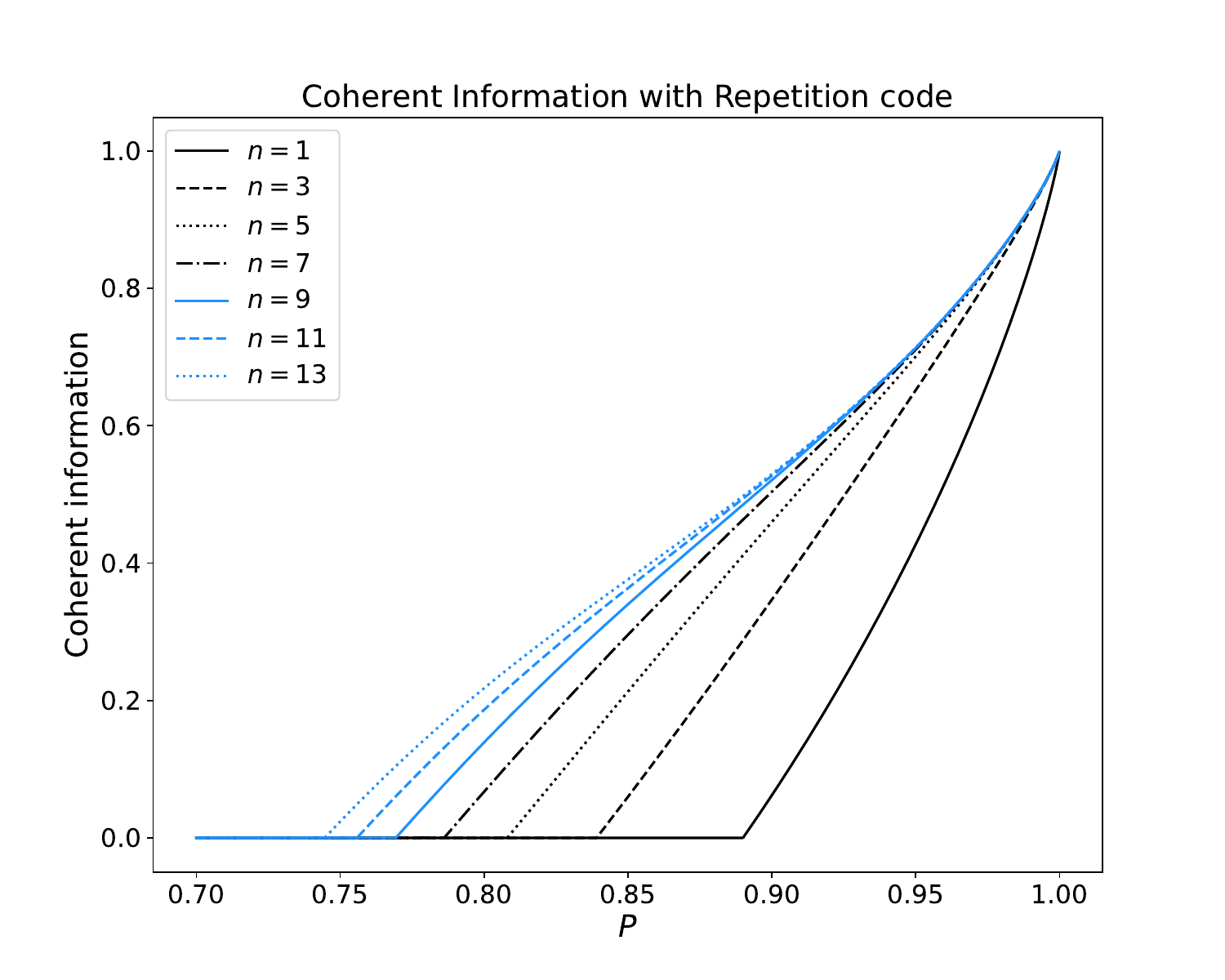}
  \caption{This plot gives the coherent information  of state where one qubit is held with Alice and $n$ qubits are with Bob. The initialization errors are on all qubits.}
  \label{fig:rep_code_2}
\end{subfigure}
\caption{The coherent information obtained with the repetition code approach outlined in the main text.}
\label{fig:coh_info}
\end{figure}

We can now plot the coherent information for the error-corrected state for various values of odd $n$ as done in Figure~\ref{fig:rep_code_1}. As observed in the plot, the state becomes more resilient to the depolarizing noise as $n$ increases.

We can, following the strategy given above, analyze the case in which all the qubits undergo the initialization error. This yields the graph given in Figure~\ref{fig:rep_code_2}. This plot also shows that on increasing the number of qubits on Bob's side the resultant state becomes more resilient towards decoding errors. However, we do observe that the coherent information does not tend to one. This can be attributed to the lack of error correction on Alice's side. As such, the coherent information would be limited by the noise incurred by Alice's qubit. We note here that the decoding strategy outlined above is not optimal. That is, the coherent information of the state obtained after the strategy is less than that of the initial state. 

\section{The case of $K_A=1$}
\label{K_A=1-section}

Here we describe how to calculate the coherent information $\sI_A$ when $K_A = 1$ since this case is rather simple and elegant. This corresponds to the rank being equal to $n_A-1$. When $n_A >3$, this will correspond to ranks greater than 2, for which the different types and patterns will be somewhat harder to keep track of. But the small value of $K_A$ allows for easy computation of the coherent information. 

For the reduced state corresponding to a given initial configuration of qubits, we obtain
\begin{equation}
\rho_{A, m_{A, 1}} = \frac{1}{2^{n_A}} \left({\mathbb I} + \prod_{l=1} ^{n_A} (-1)^{m_{A, 1}} X_{l} ^{\sJ_{A, 1l}}\right), 
\quad m_{A,1} = 0, 1\ .
\end{equation}
Let us define $q_N$ to be the probability that the product of $N$ i.i.d.~random variables $B_j \sim$ Bernoulli($P$), $1 \le j \le N$, taking $\pm 1$ values, is $+1$.
Therefore,
\begin{align}
q_N = \sum_{k=0}^{\lfloor{\frac{N}{2}\rfloor}}\binom{N}{N-2k} P^{N-2k}\left(1-p\right)^{2k}\\
=  \frac{1}{2}\left[1 + (2P-1)^N\right].
\end{align}
 
For the $K_A=1$ case with $n_A$ qubits, it is clear that
$w_0 = q_{n_A}$ and $w_1 = 1-q_{n_A}$. Therefore, $\rho_A = w_0\rho_{A, 0} + w_1\rho_{A, 1}$,
and therefore,
\be
H(\rho_A) = H_2(q_{n_A})+n_A-1.
\label{K_A=1-general-entropy-form}
\ee
Let us consider two simple examples of $K_A=1$. First an $n_A=2$ qubit subsystem example:
$\rho_{A,0} = \frac14(I + X_1X_2)$ and $\rho_{A,1} = \frac{1}{4}(I - X_1X_2)$.
Then, $w_0 = q_2 = P^2 + (1-P)^2$, and $w_1 = 2P(1-P)$,
and $H(\rho_A) = H_2(q_2)+1$,
where $H_2(x) = -x\log_2 x - (1-x)\log_2(1-x)$ is the binary entropy function.

Let us now consider an example of a three-qubit subsystem i.e.~$n_A= 3$. We then obtain
$\rho_{A,0} = \frac{1}{8}(I + X_1X_2X_4)$ and $\rho_{A,1} = \frac{1}{4}(I - X_1X_2X_4)$.
Now, $w_0 = q_3 = P^3 + 3P(1-P)^2$
and $w_1 = 1-w_0$, such that $H(\rho_A) = H_2(q_3)+2$.

\section{The case of $K_A=2$}
\label{K_A=2-section}
Here we will deal with the case of the rank being equal to $n_A-2$. As with the $k_A=1$ case, this will correspond to higher ranks when $n_A$ gets larger---computing the coherent information due to the small $K_A$ value may still be feasible.

The density operator for subsystem $A$ for a given configuration of the initial qubits for the $K_A=2$ case simplifies to 
\begin{align}
\rho_{A, m_{A,1}, m_{A,2}} = \frac{1}{2^{n_A}} &\left({\mathbb I} 
+\prod_{i=1}^{n_A} 
(-1)^{m_{A,1}} X_{i} ^{\sJ_{A, 1i}}\right)\nonumber\\
\times &\left({\mathbb I} + \prod_{j=1}^{n_A} (-1)^{m_{A,2}}X_{j} ^{\sJ_{A,2j}}\right)\ .
\end{align}

Let us say that there are $r$ and $s$ non-zero entries in $\sJ_{A,1i}$ and $\sJ_{A,2j}$, respectively,
and $t$ of the indices in $\sJ_{A,1i}$ and $\sJ_{A,2j}$ are common.
Then, we have $0 \le t \le {\min}(r,s)$,
and the total number of subsystem $A$ qubits involved $n_A = r+s-t$. Note that all these corner cases are possible: $t=r < s$, or $t=0$, or $t=s < r$. 

It is simple to argue that the weights $w_{\vec m_A}$, ${\vec m_A} = (m_{A,1},m_{A,2})$ evaluate to:
\bea
w_{00} &=& q_t q_{r-t}q_{s-t} + (1-q_t)(1-q_{r-t})(1-q_{s-t})\nonumber,\\
w_{01} &=& q_t q_{r-t}(1-q_{s-t}) + (1-q_t)(1-q_{r-t}) q_{s-t}\nonumber,\\
w_{10} &=& q_t (1-q_{r-t}) q_{s-t} + (1-q_t) q_{r-t} (1-q_{s-t})\nonumber,\\
w_{11} &=& q_t (1-q_{r-t})(1-q_{s-t}) + (1-q_t) q_{r-t} q_{s-t}\nonumber.
\label{eq:weights}
\eea

Let us consider an example involving four qubits for a reduced subsystem, with:
\begin{align}
\rho_{A, m_{A,1}, m_{A,2}} = \frac{1}{16} \left({\mathbb I} + (-1)^{m_{A,1}}X_1X_2X_3\right)\nonumber\\\left({\mathbb I} + (-1)^{m_{A,2}}X_1X_2X_4\right)\ \nonumber .
\end{align}
Since $n_A=4$, $r=3$, $s=3$, $t=2$, we have $q_t = P^2 + (1-P)^2$, $q_{r-t} = P$, and $q_{s-t} = P$. Therefore,
\begin{align}
w_{00} &= (P^2 + (1-P)^2)P^2 + 2P(1-P)(1-P)^2,\\
w_{01} &= (P^2 + (1-P)^2)P(1-P) + 2P^2(1-P)^2,\\
w_{10} &= (P^2 + (1-P)^2)P(1-P) + 2P^2(1-P)^2,\\
w_{11} &= (P^2 + (1-P)^2)(1-P)^2 + 2P^3(1-P).
\end{align}
The entropy $H(\rho_A) = H({\vec w})+n_A-K = H({\vec w})+2$.

Let us take another $K_A=2$ example:
\begin{align}
\rho_{A, m_{A,1}, m_{A,2}} = \frac{1}{16} \left({\mathbb I} + (-1)^{m_{A,1}}X_1X_2X_3\right)\nonumber\\\left({\mathbb I} + (-1)^{m_{A,2}}X_3X_4\right)\nonumber \ .
\end{align}
Here, $n_A=4$, $r=3$, $s=2$, $t=1$. We thus have $q_t = P$, $q_{r-t}=P^2 + (1-P)^2$, $q_{s-t}=P$. We can evaluate the weights $w_{\vec m_a}$ using Eqs.~\ref{eq:weights}, and $H(\rho_A)= H({\vec w})+2$.

\section{Entropy derivation for type 2}
\label{sec:subsystem_entropy_calc}
Here we will derive the subsystem entropy for a rank 2 type 2 setting. In this type, the row vectors of the biadjacency matrix $G_{AB}$ equal one of three distinct vectors.
Secondly, these three vectors satisfy the property that their binary sum gives the zero vector. Then, within each set of identical rows, we can take the first one and add it to all the others. This eliminates all of them except the first one in the set. This way, we are left with just one non-zero row of each type. But since the sum of all three of them is zero, we can eliminate one of them by adding to it the other two. We are thus left with two non-zero rows that are linearly independent.

To work out the coherent information $\sI_A$, let us say the first $n_1$ rows belong to set 1, the next $n_2$ to set 2, and the last $n_3$ rows to set 3.
Now, the full state is again a sum of several pure states, each of which corresponds to an initial configuration of the qubits (i.e.~some being initialized in $|+\ra$ and others in $|-\ra$). 
The reduced density matrix for a subsystem will thus get contributions from each such pure state.

Each such contribution has the following form,
\begin{align}
\rho_A &= \sum_{\vec m_A} w_{\vec m_A} \frac{1}{2^{n_A}}
\left[1+ (-1)^{m_{A, 1}} X_1 X_{n_1+1} X_{n_1+n_2+1}\right]\nn\\
& \times \prod_{i=2}^{n_1} \left[1+ (-1)^{m_{A, i}} X_1 X_i)\right] \nn \\
&\times \prod_{j=2}^{n_2} \left[1+ (-1)^{m_{A, n_1+j-1}} X_{n_1+1} X_{n_1+j})\right]\nn\\
&\times \prod_{k=2}^{n_3} \left[1+ (-1)^{m_{A, n_1+n_2+k-2}} X_{n_2+1} X_{n_1+n_2+k})\right], \nn \\
& m_{A, l} = 0, 1 \quad \forall l, \quad l=1\ldots n_A-2\\
& n_A=n_1+n_2+n_3.
\end{align}
Here, $\prod_{i=2}^{n_1} \left[1+ (-1)^{m_{A, i}} X_1 X_i)\right]$
corresponds to the first row of set 1 in the biadjacency matrix being added to all the other rows of set 1 in order to eliminate them. This leaves row 1 as the only uneliminated row in set 1.
Likewise, $\times \prod_{j=2}^{n_2} \left[1+ (-1)^{m_{A, n_1+j-1}} X_{n_1+1} X_{n_1+j})\right]$
and
$\prod_{k=2}^{n_3} \left[1+ (-1)^{m_{A, n_1+n_2+k-2}} X_{n_2+1} X_{n_1+n_2+k})\right]$
correspond to the elimination of all but one of the rows of sets 2 and 3 by adding to them the first row of set 2 or 3, respectively.
The first bracket i.e. $(1+ (-1)^{m_{A, 1}} X_1 X_{n_1+1} X_{n_1+n_2+1})$
corresponds to the sum of the first row from each of the three sets.

Thus, unlike type 1, here the brackets share some qubits in common.
To do the combinatorics to obtain the coefficients $w_{\vec m_A}$, first note that the first bracket has a positive sign if all three or only one of the qubits in it start in $|+\rangle$. If none or two of them start in $|-\rangle$, then it has a negative sign.

For any qubit in set 1 starting in the same initial state as qubit 1, the corresponding bracket gets a plus sign, and any qubit starting in the opposite state gives a negative sign.
The same relationship holds between the first qubit of set 2 and the other qubits of the same set, and likewise for the first qubit of set three and the rest.
We can thus write the probability coefficient of the first bracket, and multiply it by appropriate powers of $P$ and $1-P$ for the other brackets. The appropriate powers depend on which initial states are the same and which are not.

 As an example, with the first three qubits initialized in $|+\rangle$, we get $P^3$.
Now, if $n_{1+}$ of the set $1$ brackets have plus signs,
then we need $n_{1+}$ set $1$ qubits starting in $|+\ra$ and $n_1-1-n_{1+}$ in $|-\ra$, giving
$P^{n_{1+}} (1-P)^{n_1-1 -n_{1+}}$.
The same holds for sets $2$ and $3$ for which say $n_{2+}$ and $n_{3+}$ qubits are initially in $|+\ra$.
This way, we get an overall coefficient 
$P^{3 +n_{1+} +n_{2+} +n_{3+}} 
(1-P)^{n_1-1-n_{1+} +n_2-1-n_{2+} +n_3-1-n_{3+}}
= P^{3+n_{1+} +n_{2+} +n_{3+}} 
(1-P)^{n_A -3-n_{1+} -n_{2+} -n_{3+}}$,
where we have recalled that $n_A = n_1+n_2+n_3$.
Next we take a different initial configuration of the first three qubits giving the same signs for all the brackets, say one where only the first qubit starts in $|+\ra$, giving a factor of $P(1-P)^2$ instead of $P^3$. This again gives a plus sign for the first bracket, but now $P$ and $(1-P)$ switch places for set $2$ and $3$ brackets due to the second and third qubits starting in $|-\ra$. We thus write down all such configurations giving the same bracket signs and add over them.
Continuing this way, we obtain the probability coefficients
\begin{align}
&w_{\vec m_A, n_1, n_2, n_3, +, n_{1+}, n_{2+}, n_{3+}}\\
&=P^{3+n_{1+} +n_{2+} +n_{3+}} 
(1-P)^{n_A-3 -n_{1+} -n_{2+} -n_{3+}} \nn \\
&+ P^{n_{1+} +n_2-n_{2+} +n_3-n_{3+} -1}
(1-P)^{1 +n_1 -n_{1+} +n_{2+}+n_{3+}} \nn \\
&+ P^{n_1-n_{1+} +n_{2+} +n_3-n_{3+} -1}
(1-P)^{1 +n_{1+} +n_2-n_{2+} +n_{3+}} \nn \\
&+ P^{n_1-n_{1+} +n_2-n_{2+} +n_{3+} -1}
(1-P)^{1 +n_{1+} +n_{2+} +n_3-n_{3+}},
\end{align}
for configurations with a plus sign in the first bracket, $n_{1+}$, $n_{2+}$ and $n_{3+}$ plus signs in the set $1$, $2$ and $3$ brackets.
Likewise, we get
\begin{align}
&w_{\vec m_A, n_1, n_2, n_3, -, n_{1+}, n_{2+}, n_{3+}}\\
&=(1-P)^{3+n_{1+} +n_{2+} +n_{3+}} P^{n_1-n_{1+} +n_2-n_{2+} +n_3-n_{3+} -3} \nn \\
&+ (1-P)^{n_{1+} +n_2-n_{2+} +n_3-n_{3+} -1}
P^{1 +n_1 -n_{1+} +n_{2+}+n_{3+}} \nn \\
&+ (1-P)^{n_1-n_{1+} +n_{2+} +n_3-n_{3+} -1}
P^{1 +n_{1+} +n_2-n_{2+} +n_{3+}} \nn \\
&+ (1-P)^{n_1-n_{1+} +n_2-n_{2+} +n_{3+} -1}
P^{1 +n_{1+} +n_{2+} +n_3-n_{3+}}\ ,
\end{align}
for configurations with a minus sign in the first bracket, $n_{1+}$, $n_{2+}$ and $n_{3+}$ plus signs in the set $1$, $2$ and $3$ brackets.

To obtain the overall eigenvalues of the reduced density matrix, recall that for each sign configuration of the brackets obtained from a pure graph state, we obtain $2^{n_A -K_A}$ eigenstates with eigenvalues $\frac{1}{2^{n_A -K_A}}$.
With a rank 2 biadjacency matrix $K_A= n_A-2$,
this means 4 eigenstates with eigenvalues $\frac{1}{4}$.
Putting these pieces together, we obtain the eigenvalues
$\frac{1}{4}w_{\vec m_A, n_1, n_2, n_3, +, n_{1+}, n_{2+}, n_{3+}}$
and $\frac{1}{4}w_{\vec m_A, n_1, n_2, n_3, -, n_{1+}, n_{2+}, n_{3+}}$.
For the degeneracies, we count how many configurations have the same numbers of plus and minus signs for sets $1$, $2$ and $3$, and this is given by the product of the three corresponding binomial coefficients.
Combining this with the eigenvalue multiplicity of 4 for each given sign configuration, this gives the degeneracies
$4\binom{n_1-1}{n_{1+}}
\binom{n_2-1}{n_{2+}}
\binom{n_3-1}{n_{3+}}$.
And after some simplification, these give the Von Neumann entropy (\ref{subsystem-entropy-type2-rank2}) for a type 2 rank 2 subsystem of a noisy graph state.

\bibliographystyle{unsrtnat}
\bibliography{main}
\end{document}